\documentclass[11pt, a4paper]{article}
\usepackage{jheppub}
\usepackage{graphicx}
\usepackage{epsfig,amsmath,amsthm}
\usepackage{epstopdf}

\usepackage{bbold}

\newcommand{\be}{\begin{equation}}
\newcommand{\ee}{\end{equation}}
\newcommand{\bea}{\begin{eqnarray}}
\newcommand{\eea}{\end{eqnarray}}
\def\bse{\begin{subequations}}
\def\ese{\end{subequations}}

\def\IZ{\relax\ifmmode\hbox{Z\kern-.4em Z}\else{Z\kern-.4em Z}\fi}

\newcommand{\non}{\nonumber \\}
\newcommand\tensor[1]{\overset{{}_{\leftarrow\!\!\!\!\rightarrow}}{#1}}

\def\half{\frac{1}{2}} 

\def\del{{\partial}}

 \def\hC{\hat{C}} 
 
\def\hc{{\hat c}} 

\def\hd{{\hat d}} 
 
\def\hS{{\hat S}}

\def\cL{{\cal L}} 

\def\al{\alpha} \def\bt{\beta}

\def\gm{\gamma}  \def\eps{\epsilon}

\def\bi{\begin{itemize}} \def\ei{\end{itemize}}

\def\({\left(} \def\){\right)}
\def\[{\left[} \def\]{\right]}

\def\w{\omega}
\def\Om{\Omega}
\def\d{\partial}
\def\sumint{\int\!\!\!\!\!\!\!\!\sum_{\!\!\!\!\!\!\!}}
\def\sumintexplicit{\int\!\!\!\!\!\!\!\!\!\sum_{\!\!\!\!L,\w}}
\def\Lapd{\Delta_{\hd+1}}

\def\Omd{\Omega_{\hat{d}+1}}
\def\dOmd{d\Omd}
\def\Gm{\Gamma}
\def\elld{_{\ell,\hd}}
\def\Nld{N\elld}
\def\Mld{M\elld}
\def\Valph{{\aleph}}
\def\AlOm{{\aleph \, \Omega}}
\def\albt{\aleph\beth}
\def\Talbt{\aleph\beth}

\def\tldja{\tilde{j}_{\ell+\hd/2}}

\def\tldhap{\tilde{h}^+_{\ell+\hd/2}}

\def\tldhapm{\tilde{h}^\pm_{\ell+\hd/2}}
\def\altld{\tilde{\alpha}}
\def\bttld{\tilde{\beta}}
\def\gmtld{\tilde{\gamma}}
\def\IL{\mathfrak{L}}
\def\fieldh{\mathfrak{h}}		
\def\ergo{\Longrightarrow}
\def\Ms{M_S}
\def\Mv{M_V}

\def\detg{\sqrt{\!-g}}

\def \TensorNorm{{\hd^2 c_s (c_s\!-\!\hd)}}

\def \Sh{S_{EH}}
\def \Sih{S_{mat}}
\def\sourceT{\mathcal{T}}

\def\popG{\textsf{\bf G}_d}
\def\ourG{G_d}
\def\ShPreF {\frac{1}{2(\hd\!+\!1) \Omd \ourG}}
\def\ShPreFhalf {\frac{1}{4(\hd\!+\!1) \Omd \ourG}}
\def\ShPreFquarter {\frac{1}{8(\hd\!+\!1) \Omd \ourG}}
\def\ShPreFCanon{\frac{\Nld}{8(\hd\!+\!1) \ourG}}
\def\Celd {C\elld^\eps}

\definecolor{orange}{rgb}{1,0.5,0}
\definecolor{turqoise}{rgb}{0,0.5,0.5}
\definecolor{purple}{rgb}{0.4,0,0.4}
\definecolor{grey}{rgb}{0.7,0.7,0.7}
\definecolor{myyellow}{rgb}{0.7,0.7,0}
\definecolor{mygreen}{rgb}{0,0.5,0}

\def \blue{\textcolor{black}}

\def \purple{\textcolor{black}}

\newcommand{\VERBOSE}[1]{}		\newcommand{\VERBOSEE}[1]{}

\VERBOSE{

\def \blue{\textcolor{blue}}

\def \purple{\textcolor{purple}}

}

\title{Gravitational radiation-reaction in arbitrary dimension}

\author{Ofek Birnholtz and Shahar Hadar\\
{\it Racah Institute of Physics, Hebrew University, Jerusalem 91904, Israel} \\
{\tt ofek.birnholtz@mail.huji.ac.il}, {\tt shaharhadar@phys.huji.ac.il}
}

\abstract{We use effective field theory tools to study non-conservative effects in the gravitational 2-body problem in general spacetime dimension.
Using the classical version of the Closed Time Path formalism, we treat both the radiative gravitational field and its dynamical sources within a single action principle.
New results include the radiation-reaction effective action in arbitrary dimensions to leading and +1PN orders, as well as the generalized Quadrupole formula to order +1PN.}

\begin{document}
\maketitle

\section{Introduction}
\label{section:Intro}
In \cite{PaperI} an effective field theory (EFT) formalism for the simultaneous, action-level treatment of radiation and radiation-reaction (RR) effects was developed (see also \cite{PaperII} for a pedagogical introduction).
The EFT was explicitly constructed for systems of localized objects coupled to scalar, electromagnetic (EM) and gravitational fields.
The method was generalized to arbitrary spacetime dimensions in the scalar and EM cases \cite{PaperIII}.
In particular, for a single point particle, it yields the higher-dimensional analogue of the Abraham-Lorentz-Dirac (ALD) Self-Force (SF).
In this article we complete the generalization for \emph{General Relativistic} (GR) post-Newtonian (PN) systems in arbitrary dimension by applying our method and constructing the associated EFT.
The GR case is more involved than the (free) scalar and EM cases in two main respects: first, the higher spin of the gravitational field which complicates the construction of gauge-invariant fields and corresponding sources, both central features of our formalism; and second, the intrinsic non-linearity of GR. 

The PN approximation of the gravitational 2-body problem in 4 spacetime dimensions has been studied, over the past few decades, to very high accuracy (for a wide review of the field see \cite{BlanchetRev}).
The state of the art is the +4PN (corrected to order $v^{8}$) effective action, recently completed in \cite{Damour:2014jta}.
Since Goldberger and Rothstein's groundbreaking paper \cite{Goldberger:2004jt} much progress was made on the EFT approach to the PN binary problem (including \cite{GoldbergerRothstein2, GoldbergerRoss, GalleyTiglio, FoffaSturani4PNa, NRG, PortoSpin2005, PortoRothstein2006,GalleyLeibovich}); for a more detailed review section see \cite{PaperI}.
Less attention has been given to gravitational radiation and radiation-reaction in spacetime dimensions other than $4$\footnote
{
For earlier work on a fully relativistic treatment of SF for fields of various spin, see \cite{Mironov:2007nk, Mironov:2007mv, Galtsov:2007zz, Galtsov:2010cz}.
}; these will be the focus of this paper.
Although not directly relevant for astrophysical gravitating binaries, we feel that studying this rather fundamental problem is well-motivated both because an understanding of a system's behaviour in general dimension complements and enhances its understanding in $4d$ (see for example \cite{BlanchetGenD, HighLowDim, Emparan}), and because higher-dimensional gravitational scenarios frequently emerge in theoretical physics, for example in string theory.
Gravitational radiation in higher dimensions was treated in \cite{Cardoso:2002pa, Cardoso:2003jf, Cardoso:2008gn}, including in EFT methods.
Specifically, the leading-order expression for gravitational-wave emission in higher dimensions, i.e. the equivalent of Einstein's Quadrupole formula \cite{GWEinstein}, was computed.
Eq. (4.25) of \cite{Cardoso:2008gn} (also (38) \& (39) of \cite{Cardoso:2002pa}) gives the energy output in gravitational quadrupole radiation in $d$ dimensions as
\be
\frac{dE}{d\w} = \popG \frac{2^{2-d}\pi^{-(d-5)/2}d(d-3)}
	{(d-2)(d+1)\Gm\[ \frac{d-1}{2}\]}
	\w^{d+2} \left|	Q_{ij}(\w) \right|^2 \, \, .
\label{Cardoso even dimension quadrupole}
\ee

In this paper we develop an effective action for the radiation reaction of gravitation in general dimension (\ref{S eff leading}).
From this action we derive both the self-force and the dissipated power, up to subleading (+1PN) order.
At the leading order, the energy output we find (\ref{power w},\ref{power t}) matches exactly the result (\ref{Cardoso even dimension quadrupole}).
As an additional test of our results at +1PN, we substituted $d=4$ and compared with \cite{PaperI, RossMultipoles}, finding a match.

It is important to stress a relevant qualitative difference between $d\!=\!4$ and $d\!>\!4$ spacetime dimensions.
For PN astrophysical binaries RR is weak, hence it must influence the system for a long time for its effect to be significant; this is the case for bound orbits.
In higher dimensions, it is well known that there are no stable bound gravitational orbits for a binary system of gravitating compact objects\footnote{
For point particles, for example, there exist unstable circular orbits.
}.
Nevertheless, high-dimensional systems can be stabilized by non-gravitational forces (in particular short-range repulsive forces); RR may also have a substantial effect for trajectories close to the unstable circular orbit, as well as for systems with more than 2 bodies.
As we are neither specifying the sources' trajectories nor trying to solve for them, the question of binding (and its mechanisms) do not affect the results of this paper.

\subsection{Method}
\begin{figure}[t!]
\centering \noindent
\includegraphics[width=15cm]{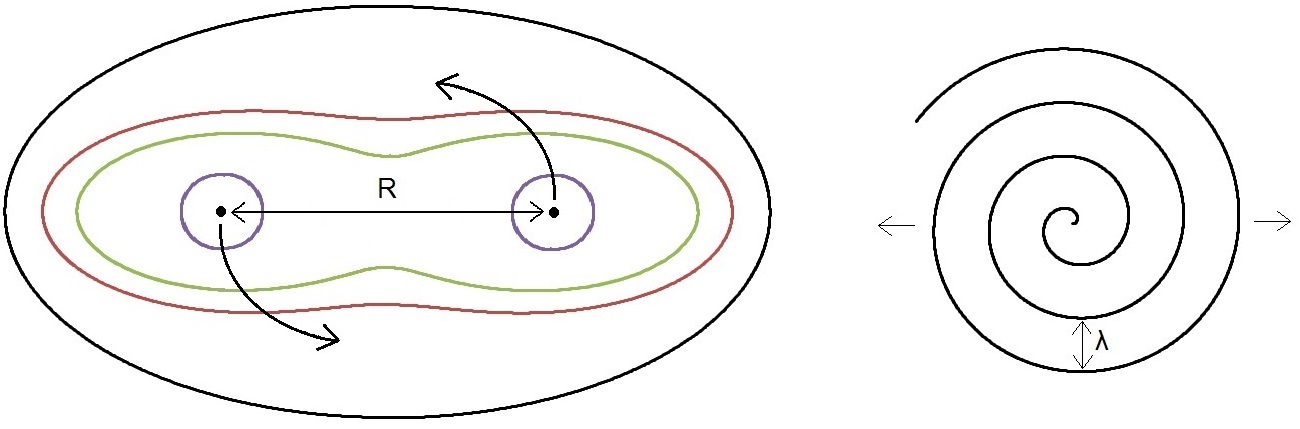}
\caption[]{The two relevant zones: the system zone is on the left, with a typical stationary-like field configuration. The radiation zone with its typical out-spiraling waves is on the right.}
 \label{fig:zones}
\end{figure}

In this work we treat the gravitational 2-body problem in arbitrary spacetime dimension, and calculate explicitly its radiation source multipoles, the RR effective action and important physical quantities derivable from it: the outgoing radiation, dissipated energy and RR force acting on the system.
Our method \cite{PaperI} divides the problem into 2 zones, the system zone and the radiation zone (fig.\ref{fig:zones}), each with different enhanced symmetries.
The system zone enjoys approximate stationarity (time independence) since by assumption all velocities are non-relativistic; we thus use the ``Non-Relativistic Gravitational'' (NRG) fields \cite{NRG} to describe it.
The radiation zone enjoys an approximate spherical symmetry, as from its point of view the system has shrunk to a point.
Hence we use \emph{gauge-invariant} spherical field variables \cite{AsninKol, KolAuxiliary, Thorne:1980ru,Kodama:2003jz,Ishibashi:2003ap} as in \cite{PaperI} and unlike the plane-wave decomposition used in most previous EFT works \cite{GoldbergerRothstein2,GoldbergerRoss, GalleyTiglio,GalleyLeibovich, FoffaSturani4PNa}).
An important ingredient of our method is the matching of these two (system and radiation) zones at the level of the action, using `2-way multipoles' \cite{PaperI}.
These are degrees of freedom we introduce - `integrate in' - to couple the two zones.
From the radiation zone point of view we think of them as sources situated at the origin while in the system zone view they reside at infinity.
At the end of the day we integrate out (eliminate) all the other degrees of freedom in the problem and remain with an effective action which is a function of these multipoles precisely.
Schwinger-Keldysh field doubling \cite{CTP}, which was beautifully adapted to the classical context in general in \cite{GalleyNonConservative} and was used in studies of the binary problem in \cite{PaperI,GalleyLeibovich}, is an essential ingredient since it allows the derivation of dissipative effects from action principles.
As we have shown in \cite{PaperIII} and is further exemplified here, a clear advantage of our formalism is its ability to naturally extend to general spacetime dimensions.

\subsection{Conventions and nomenclature}
\label{Conventions and nomenclature}
We follow the conventions of \cite{PaperIII}.
Thus the flat d-dimensional spacetime metric $\eta_{\mu\nu}$'s signature is mostly plus, and also $D:=d-1$, $\hat{d}:= d-3$. Regarding the $\hat{d}\!+\!1$ dimensional unit sphere we define $\Omd$ to be its area; on it $g_{\Om\Om'}$ is the metric, $D_\Om$ is the covariant derivative\footnote
{
At some points we will be interested in the covariant derivative with respect to an angular variable $\Om$ in the full $d$-dimensional space, and we will mark it $D^d_\Om$.
$D_\Om$ will be reserved for the more common case, of variation restricted to the $\Omd$ sphere.
Likewise the full $d$-dimensional metric, when it relates spherical coordinates, will be marked $g^d_{\Om\Om'}$.
We note that $g^d_{\Om\Om'}=r^2 g_{\Om\Om'}$, where $r$ is the radial coordinate.
}, 
and $\Lapd=D_\Om D^\Om$ is the Laplace-Beltrami operator.
The eigenfunctions of this operator on the $\hat{d}\!+\!1$ unit sphere will be given by various multipoles enumerated by an order $\ell$; in treating their eigenvalues we shall make use of $c_s=\ell(\ell+\hd)$ and $\hc_s=c_s-(\hd\!+\!1)$.
We shall designate by $\eps$ the sector (scalar, vector, tensor) of different multipoles, with either $\eps\in\{S,V,T\}$ or $\eps\in\{0,1,2\}$ respectively.

Lower case Greek letters stand for spacetime indices $\{0..D\}$, lower case Latin letters for spatial indices $\{1..D\}$, upper case Greek letters for indices on the sphere $\{1..(\hd+1)\}$, Hebrew letters ($\aleph,\beth$) for different vectorial and tensorial multipoles on the sphere, and upper case Latin letters for spatial multi-indices, i.e. $I \equiv I_\ell :=(i_\ell \dots i_\ell)$ where each $i_k\in\{1..D\}$ is an ordinary spatial index, and $\ell$ is the number of indices.
We use the multi-index summation convention with implied $\frac{1}{\ell!}$,
\be
P_I\, Q_I := \sum_\ell P_{I_\ell}\, Q_{I_\ell} := \sum_\ell \frac{1}{\ell!} P_{i_1 \dots i_\ell}\, Q_{i_1 \dots i_\ell} \, \, ,
\ee
as well as the multi-index delta function defined through $\delta_{I_\ell J_\ell} := \ell!\, \delta_{i_1 j_1} \dots \delta_{i_\ell j_\ell}$, so that factors of $\ell!$ are accounted for automatically.

While as usual $c=1$, we specifically wish to keep the gravitational constant in $d$ dimensions, marked $\ourG$.
We choose it and the normalization for the gravitational action so that the Newtonian potential is always $-\ourG M / r^\hd$, and the action is given by (\ref{action EH}).
It is related to the commonly used constant $\popG$\footnote{
In terms of which the overall prefactor of the Einstein-Hilbert action is $\frac{1}{16 \pi \, \popG}$ in arbitrary dimension and black-hole entropy is given by $\frac{\mathcal{A}_H}{4\, \popG}$.
}
by $\ourG = \frac{8\pi}{(\hd\!+\!1) \Omd} \, \popG$.
Both definitions identify for $d=4$ \cite{EmparanReall, Cardoso:2008gn}.
The constant $\ourG$ also has units of ${L^{\hd}}/{M}$.

We use the convention that the Feynman rules are real \cite{CLEFT-caged}, as befits a non-quantum field theory.
We also denote \be \sumint \triangleq \int\!\frac{d\w}{2\pi} \sum_{L}~. \ee

\section{From Einstein's Action to Feynman Rules}

\subsection{Action for metric perturbations}
\label{action_for_pert}
In GR, the action for the metric $\tilde{g}_{\mu\nu}$ coupled to matter is given by the sum of the Einstein-Hilbert (EH) Action and a matter term:
\be
S = \int \!\! d^d x \, \sqrt{-\tilde{g}} \, \[ \ShPreF \, R + \cL_M \] ~,
\label{action EH}
\ee
where $\cL_M$ is the matter Lagrangian.
We shall calculate the generation and reaction of Gravitational Waves (GWs) propagating out to infinity in asymptotically flat space, in spherical coordinates.
The metric splits to $\tilde{g}_{\mu\nu}=g_{\mu\nu}+h_{\mu\nu}$, where $g_{\mu\nu}$ describes the background flat spacetime in spherical coordinates\footnote{
See appendix \ref{app:spherical harmonics}; henceforth, covariant derivatives are associated with this metric.
} 
and $h_{\mu\nu}$ is a perturbation on it.
The EH term is the kinetic part of the action, and describes propagation of GWs.
The matter term describes the generation of GWs from (matter) sources.
We expand the kinetic (source) action to quadratic (linear) order in $h_{\mu\nu}$ to find
\bea
\Sh &=& \ShPreFhalf \int \!\! d^d x \detg
	\[\frac{1}{2} \, h^{\al ~ \gm}_{\,[\al ;} h^{\,\bt}_{\bt];\gm}
		+ h^{\gm ~\, \bt}_{\,[\bt ;} h^{\,\al}_{\gm];\al}
	\] \! , ~~~~
\label{S_hom}
\label{action homogenous} \\
\Sih &=& \int \!\! d^d x \detg \, h_{\mu\nu}T^{\mu\nu} ~.
\label{action inhomogenous}
\eea
We use the action composed of (\ref{S_hom}) and (\ref{action inhomogenous}) to construct our \emph{far zone} action in terms of spherical variables.
As we compute corrections only up to +1PN, and as far zone nonlinearities enter only at orders higher than +1PN (see discussion in section \ref{rad zone corrections}), we need not take them into account here; this justifies expanding the action to quadratic order in $h$.
The \emph{near zone} (a.k.a ``system zone") nonlinearities \emph{do} enter at +1PN, and are treated in section \ref{source non-linearities}.

We use spherical harmonics to decompose the fields and sources to 11 families of fields, comprised of the 7 scalar families $h_{tt}$, $h_{tr}$, $h_{rr}$, $h_{tS}$, $h_{rS}$, $h_S$, $\tilde{h}_S$, the 3 vector families $h_{t\Valph}$, $h_{r\Valph}$, $h_{\Valph}$, and the single tensor family $h_{\Talbt}$, defined through (see appendix \ref{app:Spherical fields})
\bea
h_{tt} &=& \sumint h^{L\w}_{tt} n_L e^{-i\w t} \triangleq \int\!\frac{d\w}{2\pi} \sum_{L} h^{L\w}_{tt} n_L e^{-i\w t} ~~,\nonumber\\
h_{tr} &=& \sumint h_{tr}^{L\w} n_L e^{-i\w t}~~~~~~~~~~~~~~~~~~~~~~~~,~~
h_{rr} = \sumint h_{rr}^{L\w} n_L e^{-i\w t} ~~,\nonumber\\
h_{t\Om} &=& \sumint \( h_{t}^{L\w} \d_\Om n_L + h_{t\Valph}^{L \w} n^{L}_\AlOm \) e^{-i\w t} ~~,~~
h_{r\Om} = \sumint \( h_{r}^{L\w} \d_\Om n_L + h_{r\Valph}^{L \w} n^{L}_\AlOm \) e^{-i\w t} ~~,\nonumber\\
h_{\Om\Om'} &=& \sumint \[ h^{L\w}_{S} n^L_{\Om\Om'}
	+ \tilde{h}^{L\w}_S \tilde{n}^L_{\Om\Om'}
	+ h^{L\w}_\Valph n^{L}_{\AlOm\Om'}
	+ h^{L\w}_{\Talbt} n^{L}_{\albt\Om\Om'} \] e^{-i\w t}~~,
\label{decomposition of GR fields}
\eea
where we use the scalar multipoles $n^L$, the vector multipoles $n^{L}_\AlOm$
and the tensor multipoles $n^{L}_{\aleph\aleph'\Om\Om'}$ described in appendix \ref{app:Spherical fields}.
They are dimensionless and depend only on the angular coordinates.
We shall at times omit $L,\w$ indices for brevity.

Substituting the new fields (\ref{decomposition of GR fields}) into the homogenous action (\ref{S_hom}) and using the definitions (\ref{solid harmonics}), derivative relations (\ref{derivatives of h}-\ref{Laplacians3}) and normalization relations (\ref{sphercal harmonics normalization}) we find the homogenous action decomposes into 3 independent sectors, namely spherical scalar, vector and tensor.
Following \cite{AsninKol} we solve for the algebraic fields (one vector and three scalars) and reduce further over the gauge degrees of freedom (again one free function in the vector sector and three in the scalar).
We are thus left with 3 types of gauge-invariant master fields, one type in each sector \cite{Gerlach:1979rw, ReggeWheeler, Zerilli:1970se, Moncrief:1974am, Cunningham:1978zfa, KodamaIshibashi, MartelPoisson}; we call them $\fieldh^{L\w}$, $\fieldh^{L\w}_{\Valph}$, $\fieldh^{L\w}_{\Talbt}$.
These appear in the action in similar forms (see appendix \ref{app:Homogenous action in spherical fields}, in particular (\ref{Tensor Action Master Homogenous}, \ref{Vector Action Master Homogenous}, \ref{Scalar Action Master Homogenous})):
\bea
\Sh
&=& \sumint \ShPreFCanon
	\int \!\! dr \, r^{2\ell+\hd+1} \!
	\[
		\frac{4 (\hd\!+\!1) (\ell\!-\!1) \ell }{\hd (\ell\!+\!\hd\!+\!1) (\ell\!+\!\hd)} \fieldh^{*} \IL \, \fieldh
		+\frac{4(\ell\!-\!1) (\ell\!+\!\hd)}{(\ell\!+\!\hd\!+\!1) \ell} \fieldh^{*}_{\Valph} \IL \, \fieldh_{\Valph}
	\right. \nonumber\\ \nonumber\\ && \left. ~~~~~~~~~~~~~~~~~~~~~~~~~~~~~~~~~~~~~
		+\TensorNorm \, \fieldh^{*}_{\Talbt} \IL \, \fieldh_{\Talbt}
	\],~~~
\label{Homogenous Action by sectors}
\\
\IL &=&	\(	\w^2
			+ \d_r^2
			+\frac{2\ell+\hd+1}{r}\d_r
		\),
\label{Master wave operator}
\eea
where $\Nld \!=\! \frac{\Gm(1+\hd/2)} {2^\ell\,\Gm(\ell+1+\hd/2)} \!=\! \frac{\hd!!}{(2\ell+\hd)!!}$.
We note that for $d=4$ the scalar and vector expressions match correspondingly eq. (3.80) of \cite{PaperI}; there is no tensor sector in $4d$.

For the inhomogenous (source) part of the action (\ref{action inhomogenous}), we similarly decompose the sources $T^{\mu\nu}$ to $T^{tt}$, $T^{tr}$, $T^{rr}$, $T^{t}$, $T^{r}$, $T^{S}$, $\tilde{T}^{S}$, $T^{t\Valph}$, $T^{r\Valph}$, $T^\Valph$, $T^{\Talbt}$ as in (\ref{decomposition of T}).
Collecting the sources into the combinations $\sourceT$, $\sourceT^{\Valph}$, $\sourceT^{\Talbt}$ matching the gauge-invariant master fields (see appendix \ref{app:Inhomogenous action in spherical fields - Source terms}, in particular (\ref{TS}-\ref{TT})), we find the action for the interaction of the fields with their sources is given by (\ref{inhomogenous source}), and the complete action in our gauge-invariant variables is comprised of three parts
\be
S = S^S + S^V + S^T,
\label{Action S}
\ee
where
\be
S^\eps = \half ~ \sumint \int dr \[ \frac{\Nld}{\ourG R^\eps\elld} \, r^{2\ell+\hd+1}
					\fieldh^{*}_{\eps} \IL \, \fieldh^{\eps}
				- \( \fieldh^{*}_{\eps} \sourceT^{\eps} + c.c. \)
				\],
\label{Master Action}
\ee
for all sectors $\eps\in\{S,V,T\}$, and
\be
R^S\elld = \frac{\hd (\ell\!+\!\hd\!+\!1) (\ell\!+\!\hd)}{(\ell\!-\!1) \, \ell }~,~
R^V\elld = \frac{(\hd\!+\!1) (\ell\!+\!\hd\!+\!1) \ell}{2 (\ell\!-\!1) (\ell\!+\!\hd)}~,~
R^T\elld = \frac{8 (\hd\!+\!1)}{\TensorNorm}~.
\ee
Variation of (\ref{Master Action}) yields the master wave equation (compare (2.14, 3.18, 3.26) of \cite{PaperIII}, and (2.9, 3.8, 3.49, 3.57) of \cite{PaperI}\footnote{
Notice that our subscripts $\ell,\hd$ in $R\elld^\eps$ carry a different meaning than the subscript $s$ in \cite{PaperI}'s $R^\eps_s$.
The $s$ in \cite{PaperI} denotes the field type, which for gravity is always 2; we choose to emphasize the $\ell,\hd$ dependence.
}),
\be
0=\frac{\delta S}{\delta \fieldh^{L\w *}_{\eps}}
	= \frac{\Nld}{\ourG R^\eps\elld} \, r^{2\ell+\hd+1}  \( \w^2 + \d_r^2 +\frac{2\ell+\hd+1}{r}\d_r \) \fieldh^{\eps}_{L\w}
	-\sourceT^{\eps}_{L\w}~.
\label{Master Wave Equation}
\ee

\subsection{Feynman Rules - Propagators and Vertices}
\label{feynman rules}
The solutions to the homogenous part of such wave equations are composed using the origin-normalized Bessel functions $\tldja(\w r)$ and $\tldhap(\w r)$ (see appendix \ref{app:Normalizations of Bessel functions} and \cite{PaperIII}).
Hence in the language of Feynman diagrams, we can represent the retarded propagator for the various waves using the diagram
\bea
\parbox{20mm}{\includegraphics[scale=0.5]{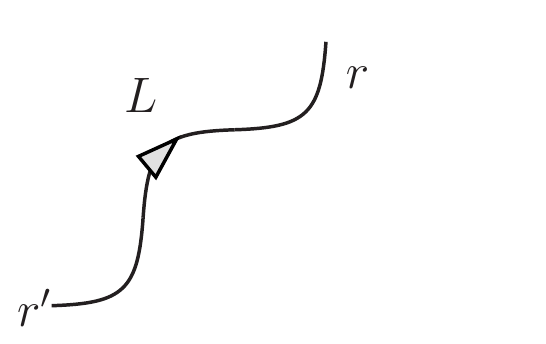}}
 \equiv G^\eps_{ret}(r',r)
= -i \w^{2\ell+\hd} ~ \ourG ~ \Celd ~ \tldja(\w r_1) ~ \tldhapm(\w r_2) \, \, ;~~~~
\label{Master Propagator}
\eea
with $r_1:=\min\{r',r\}$, $r_2:=\max\{r',r\}$, $\Celd = \Mld \, R^\eps\elld$ and
\be
\Mld = \frac{\pi}{2^{\ell+1+\hd}\Gm(1+\hd/2) \, \Gm(\ell+1+\hd/2)}~.
\label{Mld}
\ee
In particular $\Mld=\[ \hd!!(2\ell+\hd)!! \] ^{-1}$ in odd $\hd$, $\Mld=\frac{\pi}{2} \[ \hd!!(2\ell+\hd)!! \] ^{-1}$ in even $\hd$.

The vertices (source multipole terms) in the radiation zone are found by matching with the system zone according to the diagrammatic definition
\bea
- Q^\eps_{L\w} =
\parbox{65mm}{\includegraphics[scale=0.5]{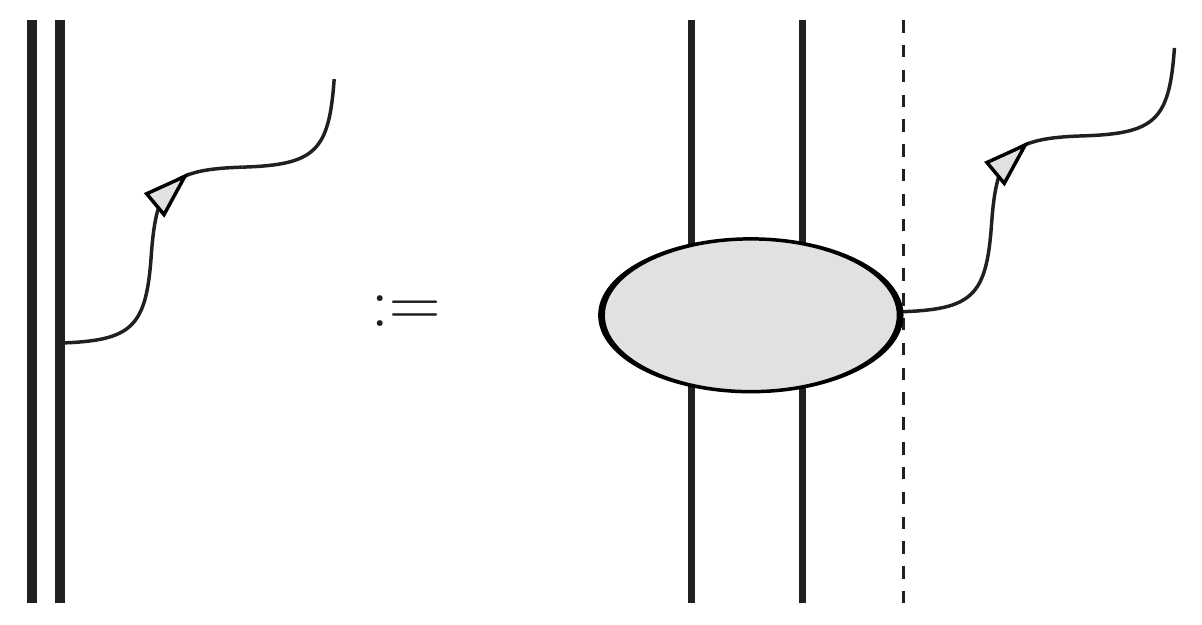}}
,
\label{vertex_definition}
\eea
where the blob on the right hand side means one should sum over all possible near-zone diagrams with an outgoing radiation leg.
From the radiation zone's point of view the sources $Q^{L \w}$ are located at the origin $r=0$, hence the radiation zone field can be written as
\bea
\fieldh_{L \w}^{\eps,EFT}(r) =
\parbox{20mm} {\includegraphics[scale=0.5]{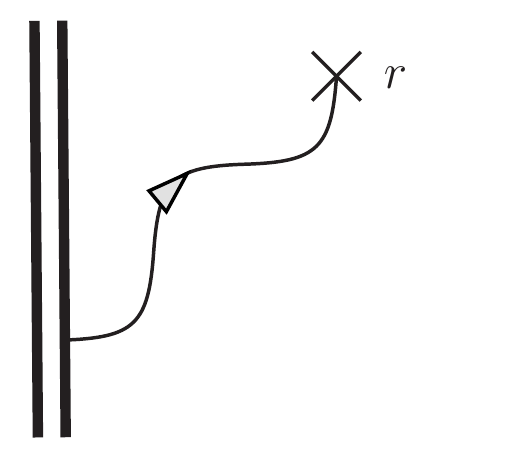}}
= - Q^{L \w}_\eps  \( -i \w^{2\ell+\hd} \, \ourG \, \Celd  \)  \tldhap(\w r) \, \, .
\label{wavefunction at radiation zone1}
\eea
On the other hand, in the full theory we may also use spherical waves to obtain the field outside the source as
\bea
\fieldh^{\eps}_{L \w}(r)
	&=&  -  \int \!\! dr' \, \sourceT^{\eps}_{L \w}(r') G^{\eps}_{ret}(r',r)  \\ 
	&=&  -  \[ \int \!\! dr' \, \sourceT^{\eps}_{L \w} (r') \tldja(\w r')  \! \]
		\( -i \w^{2\ell+\hd} \, \ourG \, \Celd  \) \tldhap(\w r). \nonumber  
\label{wavefunction at radiation zone2}
\eea
From the comparison between (\ref{wavefunction at radiation zone1},\ref{wavefunction at radiation zone2}) we read off the source multipoles as
\be
Q^{L \w}_\eps = \int \!\! dr \tldja(\w r) \sourceT^{\eps}_{L \w} (r).
\label{source Q}
\ee
Following \cite{RossMultipoles,PaperI} we call this matching process a ``zoom out balayage''\footnote{
French for ``sweeping away"; Poincar\'e coined the term (\cite{{balayage-encyc}, Balayage-Poincare}), describing the process where a charge distribution in some spatial region is ``swept away'' to the boundary of that region, leaving the potential outside unchanged.
} 
of the original source distribution $\sourceT^\eps_{\w}({\vec{r}})$ out to the multipole $Q^{L\w}_\eps$, carried out through propagation with $\tilde{j}_{\ell+\hd/2}(\w r)$.
Thus we find both the source vertex and vertex for the doubled (hatted) source,
\begin{align}
\label{Feynman Rule Vertex}
\parbox{13mm} {\includegraphics[scale=0.3]{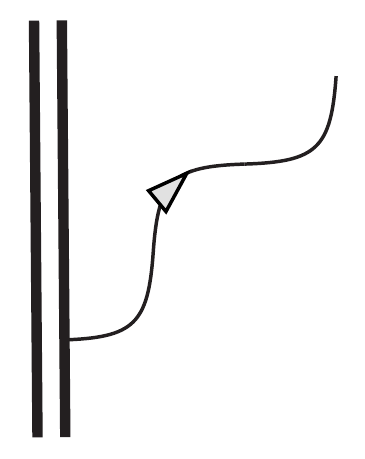}}
=-Q^{L \w}_{\eps}\,~~~~~ ,~~~
\parbox{13mm} {\includegraphics[scale=0.3]{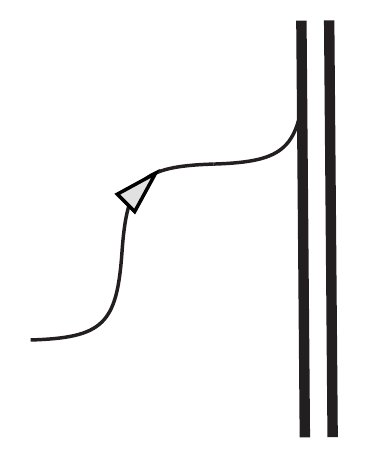}}
=-\hat{Q}^{L \w}_{\eps \,*}~~~~ .
\end{align}
At this point we first introduce field doubling and employ the classically adapted Closed-Time-Path formalism; we work in the Keldysh basis.
For a detailed description, see \cite{GalleyNonConservative,PaperI}.

It is worthy to note that the series expansions of both $\tilde{j}_{\ell+\hd/2}$ (\ref{Bessel J series2}) and of the sources $\sourceT^{\eps}_{L \w}$ (\ref{TS},\ref{TV},\ref{TT}) include only integer powers of $\w$.
This implies analyticity of the multipoles $Q(\w)$ as functions of (complex) frequency, for any $\ell$ and $\hd$, which in turn results in local-in-time expressions for the time-domain multipoles $Q(t)$ - that is, the multipoles at a certain time $t$ themselves depend on the energy momentum tensor at the same $t$ alone and do not contain tails.
The effective action $\hat{S}$ \emph{can} still contain tails - this will happen in the case of non-even dimension \cite{Balazs, Ching, Chu}.
Using the inverse transformations (\ref{invTtt}-\ref{invTT}), source terms (\ref{TS}-\ref{TT}), integration by parts and the modified Bessel equation (\ref{Modified Bessel equation2}), we find the \emph{radiation source multipoles} in the frequency domain
\bea
Q^{L\w}_S &=& \frac{(\hd\!+\!1) \ell}{\hd c_s (\ell\!+\!\hd\!+\!1)} \! \int \!\! d^D x \, x^L_{STF} \!
\[	\hd\, T^{tt}_{\w}
		\( \frac{c_s}{\hd+1}\!+\!\ell\!+\!\hd \!-\! \frac{\w^2 r^2}{\hd} +r\d_r\)
	\right. \nonumber \\ &&~~~~~~~~~~~~~~~~~~~~~~~~~~~~~~~~	   \left.
	+2 i \w r\, T^{tr}_{\w}
		\( \ell\!+\!\hd\!+\!1+r\d_r \)
	\right. \nonumber \\ &&~~~~~~~~~~~~~~~~~~~~~~~~~~~~~~~~	   \left.
	+T^{aa}_{\w}
		\( \frac{c_s}{\hd\!+\!1} \!+\! \ell \!+\! \hd + r\d_r \)
	- \w^2 r^2\, T^{rr}_{\w}
\] \tldja(\w r) ~,
\label{QSw}
\nonumber \\
Q^{L\w}_{\Valph}
\VERBOSE{	
&=& \frac{2\Nld \Omd (\ell\!+\!\hd)}{\ell\!+\!\hd\!+\!1} \!\!\! \int \!\! dr\, r^{\hd+1}
	\[	r^{1-\hd} T^{t\Valph}_{L\w} \d_r \( r^{\ell+\hd+1} \tldja(\w r) \)
		+ i\w r^{\ell+2} T^{r\Valph}_{L\w} \tldja(\w r)
	\]
\nonumber \\
}	
&=& \frac{2\eps^{(D)}_{\aleph a b k_\ell} }{\ell\!+\!\hd\!+\!1} \int\!d^D x
	\[	r^{-(\ell+\hd)} x^{bL-1} T^{ta}_{\w} \d_r \( r^{\ell+\hd+1} \tldja(\w r) \)
		+ i\w T^{ac}_{\w} x^{bcL-1} 
	 \tldja(\w r)
	\]\!,
\label{QVw}
\nonumber\\
Q^{L\w}_{\Talbt}&=& \frac{\ell(\ell\!-\!1)}{2}
	\eps^{(D)}_{\aleph a b k_{\ell}} \eps^{(D)}_{\beth a' b' k_{\ell'}} 
	\int \!\! d^D x\, x^{bb' \! L-2}  T^{aa'}_{\w} \tldja(\w r) \,
\label{QTw}~.
\eea
Fourier transforming gives the multipoles in the time domain, recorded explicitly in (\ref{QS}, \ref{QV}, \ref{QT}).
These may also be written using the $\hd$-dimensional generating time-weighted function $\delta_{\elld}(z)$ defined in \cite{PaperIII} (following the $4d$ definition \cite{Blanchet:1989ki,Damour:1990gj}), as
\be
Q^L_\eps = \int d^D x \, x^L_{STF}
	\int_{-1}^{1}dz \delta_{\elld}(z) \sourceT^\eps(\vec{r},u+z r) \, ~ .
\label{I Phi scalar multipoles using delta ell}
\ee
The multipoles represent the gravitational source scalar, vector and tensor multipole moments.
A simple test of substituting $d=4$ shows these multipoles coincide (for any $\ell$) with the scalar and vector multipoles given by \cite{RossMultipoles,PaperI} (see (3.88) and (3.89) of \cite{PaperI}); there is no tensor sector in $4d$.
In general $d$, for a system of $N$ masses at leading PN order (low velocities, weak potential), the leading multipoles (of order $\ell=2$, or quadrupole) of each sector are the mass (scalar, or electric) quadrupole, the gravitational source-current (vector, or magnetic, +1PN) quadrupole\footnote{
The factor of 2 between the gravitational source current and the mass current, follows app. A.3 of \cite{PaperI}.
}, and the mass-tensor quadrupole (+2PN), respectively:
\bea
Q^{ij}_{S} &\to& \sum^n_{A=1} m_A \( x^i x^j - \frac{1}{D}\delta^{ij}x^2\)_A~,
\label{scalar quadrupole leading}\\
Q^{ij}_{\Valph} &\to& 2 \sum^n_{A=1} m_A \( x^i \eps^{j}_{\aleph a b} x^a v^b \)_A^{TF}~,
\label{vectorr quadrupole leading}\\
Q^{ij}_{\Talbt} &\to& \sum^n_{A=1} m_A \( \eps^{i}_{\aleph a b} \, x^a v^b \eps^{j}_{\aleph c d} \, x^c v^d \)_A^{TF}~.
\label{tensor quadrupole leading}
\eea
We note again at this point that in order to go \emph{beyond} the +1PN approximation more Feynman rules, that correspond to radiation zone interactions, need to be included; see discussion in sections \ref{action_for_pert} and \ref{beyond 1PN}.

\subsection{Source Non-linearities}
\label{source non-linearities}
In order to describe the radiative multipoles of a system of masses beyond the leading orders, we prescribe how the system's $T^{\mu\nu}$ is given by the masses and their coordinates.
From the inhomogenous coupling (\ref{action inhomogenous}), $T^{\mu\nu}$ is the source for $h_{\mu\nu}$, represented as a 1-pt diagram (compare \cite{PaperI,Blanchet:1998in}),
\bea
  T^{\mu\nu}(x) := -2 \frac{\delta S_{eff}}{\delta h_{\mu\nu}(x)} \equiv -2\,
 \( \parbox{30mm}{\includegraphics[scale=0.3]{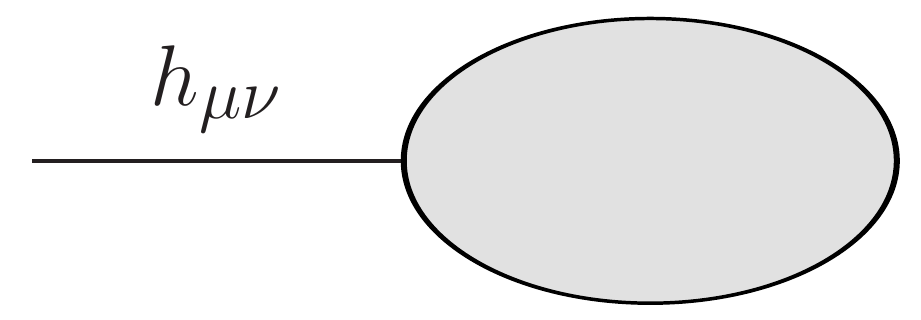}} \)~.
\label{Tmunu source}
\eea
In the system zone we describe the metric using the three NRG fields
$h_{\mu\nu} \longleftrightarrow \( \phi,\, \vec{A},\, \tensor{\sigma}\)$,
defined by
\be
ds^2 = e^{2\phi}\( dt - 2\vec{A}\cdot d\vec{x}	\)^2 -e^{-2\phi/\hd} \( \delta_{ij}+\sigma_{ij} \) dx^i dx^j.
\label{NRG fields}
\ee
This definition almost matches that of \cite{CLEFT-caged,NRG}, but includes a factor of $2$ in the defintion of the coupling of the gravitomagentic vector potential $\vec{A}$, as introduced in the $4d$ case in eq. (A10) of \cite{PaperI}.
The corresponding sources are given by
$T^{\mu\nu} \longleftrightarrow \(\rho_\phi,\, \vec{J},\, \tensor{\Sigma} \)$,
defined by
\be
\Sih = -\rho_\phi\, \phi + \vec{J} \cdot \vec{A} - \half \Sigma^{ij} \sigma_{ij} ~.
\label{NRG couplings}
\ee
This implies
\bea
T^{tt} + \frac{1}{\hd}T^{ii} &=& \rho_\phi
= -\( \parbox{30mm}{\includegraphics[scale=0.3]{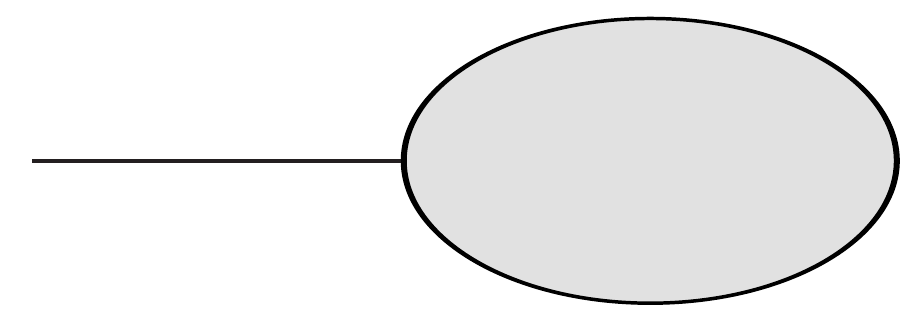}} \) \, \, ,
\non
T^{ti} &=& J^i
= \parbox{30mm}{\includegraphics[scale=0.3]{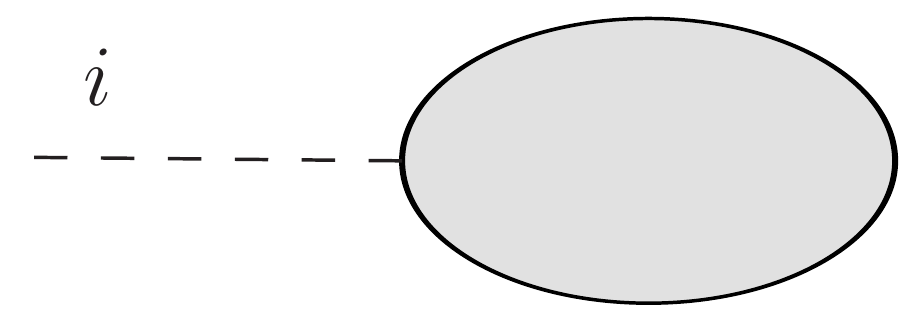}} \, \, ,
\non
T^{ij} &=& -\Sigma^{ij} =
2\, \parbox{30mm}{\includegraphics[scale=0.3]{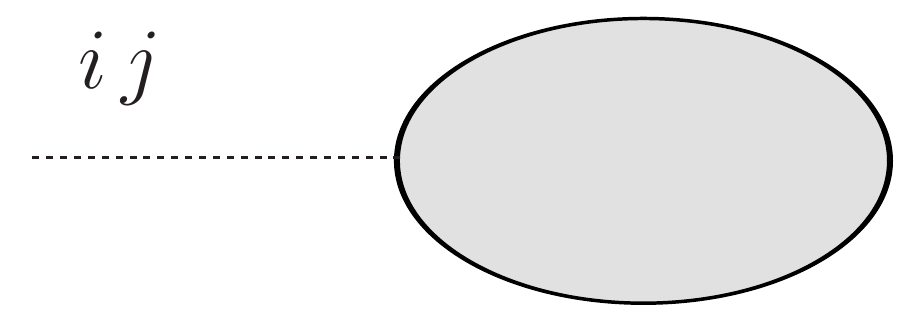}}  ~.
\label{NRG source definitions}
 \eea
We refer to $\rho_\phi$ as the gravitational mass density, to $\vec{J}$ as the gravitational source current, and to $\Sigma^{ij}$ the gravitational stress; together they comprise the NRG source fields.
We further mark $\Sigma=\Sigma^{aa}$, and note that $T^{tt}=\rho_\phi+\frac{1}{\hd}\Sigma$.
In terms of these source fields we re-write the expression for the radiation source multipoles (\ref{QS},\ref{QV},\ref{QT}) as follows 

\bea
Q^L_S &=& \frac{(\hd\!+\!1)}{\hd (\ell\!+\!\hd) (\ell\!+\!\hd\!+\!1)} \! \int \!\! d^D x \, x_{STF}^L \!
\[	\hd \( \frac{(\ell\!+\!\hd)(\ell\!+\!\hd\!+\!1)}{\hd+1} + r\d_r \) \!
		 \tldja(i r \d_t) \rho_\phi
	\right. \nonumber \\ &&~~~~~~~~~~~~~~~~~~~~~~~~~~~~~~~~~~~~~   \left.
	-2 r^{-(\ell\!+\!\hd)}\(r^{\ell\!+\!\hd\!+\!1}\tldja(i r\d_t)\)'\!\!\vec{x}\!\cdot\!\dot{\vec{J}}
	\right. \nonumber \\ &&~~~~~~~~~~~~~~~~~~~~~~~~~~~~~~~~~~~~~   \left.
	+ \tldja(i r \d_t) \( r^2 \( \ddot{\rho}_\phi + \frac{1}{\hd} \ddot{\Sigma} \)
	- x_a x_b \, \ddot{\Sigma}^{ab} \)
\] \! , \,
\label{QS NRG}	\nonumber \\ \nonumber \\
Q^L_{\Valph}&=&\frac{\eps^{(D)}_{\aleph a b k_\ell} }{\ell\!+\!\hd\!+\!1} \!\! \int\!\!d^Dx\! 
					\[ \( r^{\ell\!+\!\hd\!+\!1} \tldja(i r \d_t) \)'
						\frac{x^{bL-1}}{r^{\ell\!+\!\hd}} J^a
					  + \tldja(i r \d_t) \dot{\Sigma}^{ac} x^{bcL-1}
					\],~~~~~~~
\label{QV NRG}	\nonumber \\ \nonumber \\
Q^L_{\Talbt}
&=& -\frac{\ell(\ell\!-\!1)}{2}
		\eps^{(D)}_{\aleph a b k_{\ell}} \eps^{(D)}_{\beth a' b' k_{\ell'}} 
		\int \!\! d^D x\, x^{bb' \! L-2}  \Sigma^{aa'}(\vec{r},t) \tldja(i r \d_t) \,
\label{QT NRG}~.
\label{Q NRG}
\eea
We remark that upon substitution of $d=4$, equations (\ref{Q NRG}) reproduce equations (3.105) of \cite{PaperI} (using (\ref{Modified Bessel equation2})).
To use the NRG source fields defined in (\ref{NRG source definitions}), we expand the diagrams perturbatively in PN orders.
We now focus on the $N$-body problem of masses $\{m_A\}_{A=1}^{N}$ whose trajectories are described by $\{\vec{x}_A(t)\}_{A=1}^{N}$, and influenced by their gravitational interactions.
For such systems the gravitational mass density at leading and next to leading (+1PN) orders is found by Taylor expanding (\ref{NRG fields}) and varying w.r.t $\phi$ (using (\ref{NRG couplings})) to be
\bea
\rho_\phi (\vec{x},t) &=&
	- \( \parbox{25mm}{\includegraphics[scale=0.5]{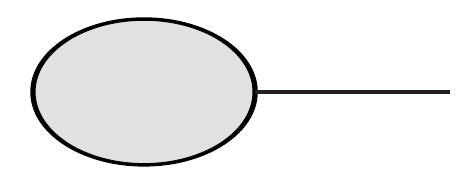}} \) =
	- \( \parbox{18mm}{\includegraphics[scale=0.5]{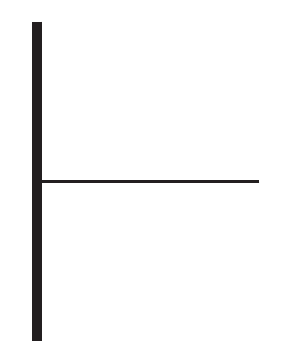}}
		+ \parbox{18mm}{\includegraphics[scale=0.5]{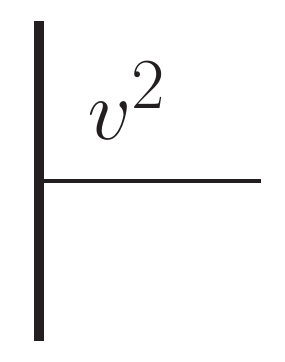}}
		+ \parbox{25mm} {\includegraphics[scale=0.5]{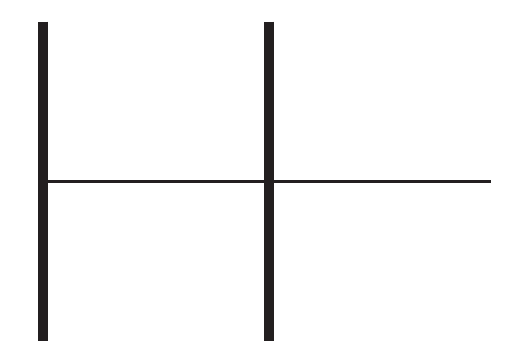}}
	  \)
\non
 &=&  \sum^n_{A=1} m_A\, \delta\(\vec{x}-\vec{x}_A (t)\) \,
		\(1
		  + \frac{\hd+2}{2 \, \hd} v_A^2(t)
		  - \sum_{B\neq A} \frac{\ourG\, m_B}{\| \vec{x}_A(t) \!-\! \vec{x}_B(t)\|^\hd}
		\) ~.
\label{source rhophi}
\eea
When other (not only gravitational) interactions are involved in maintaing the bound system (see \ref{section:Intro}), they also contribute at +1PN.
Their potential energy is to be included in the nonlinear correction (\ref{source rhophi}), similarly to its last term (or in its stead, if the gravitational potential is much weaker then the PN parameter, $\phi_{gravity}\ll v^2\sim\phi_{other}$ ).

Non-linear contributions to the additional NRG source fields $\vec{J}$ and $\Sigma^{ij}$ are to be derived in similar fashion.
As we here calculate only up to +1PN, the leading orders of $\vec{J}$ and $\Sigma^{ij}$ suffice:
\bea
\vec{J} &=& 2 \sum^n_{A=1} m_A\, \vec{v}_A\, \delta\(x-x_A\)~,
\label{source J}	\\
\Sigma^{ij} &=& \sum^n_{A=1} m_A\, v_A^i\, v_A^j\, \delta\(x-x_A\) ~.
\label{source Sigma}
\eea
These of course match (3.107,3.108) of \cite{PaperI} for $\hd=1$.

\section{Application \& results}
\label{Results}
Having set up the apparatus of Feynman rules (by sector $\eps$ and by multipole order $\ell$, in (\ref{feynman rules})), we now diagramatically construct the RR effective action \cite{PaperI, PaperII, PaperIII}.
We expand order by order in the PN expansion, i.e. in the small parameter $v^2 \sim \frac{\ourG M}{r^\hd}$, and recalling that $\w r \sim v^{1/2}$ counts as half an order.
A general contribution to this effective action is given by a vertex-propagator-vertex diagram,
\bea
\parbox{20mm}{\includegraphics[scale=0.5]{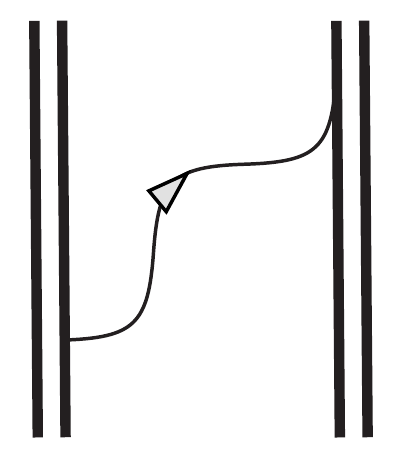}}
 \equiv Q_\eps^{L\w} G^\eps_{ret}(r=0,r'=0) \hat{Q}_{\eps}^{L\w \, *}
= -i \w^{2\ell+\hd} \, \ourG \, \Celd \, Q_\eps^{L\w} \hat{Q}_{\eps}^{L\w \, *} \, + c.c. 
\label{S eff general}
\eea
Below we calculate the RR effective action to the leading and next-to-leading orders; we then discuss higher order contributions.

\subsection{Leading Order}
\label{LO}
The only contribution at leading order (LO) is that of the scalar quadrupole ("Mass Quadrupole", $s\!=\!0,\,\ell\!=\!2$, also called E2),
\bea
S_{eff,\w}^{LO}=\parbox{20mm}{\includegraphics[scale=0.5]{ActionDiagScalar.pdf}}	
&=& -i \w^{\hd+4} \, \ourG \, C^S_{2,\hd} \, Q_S^{L_2\w} \hat{Q}_{S*}^{L_2\w} \, + c.c.
\nonumber\\
&=& - i \w^{\hd+4} \, \ourG
	\frac{\pi \, \hd (\hd\!+\!3) (\hd\!+\!2)}{2^{\hd+4}\, \Gm(3+\hd/2)\Gm(1+\hd/2)} \,
	 Q_S^{L_2\w} \hat{Q}_{S*}^{L_2\w} \, + c.c. 
\label{S eff leading}
\eea
Following the procedure of \cite{PaperI} ((2.57-2.61) \& Sec. III.C.1), we can immediately find from this expression the RR force on the system's constituent masses, the effective potential, and the energy dissipated as radiation.
In the remainder of this section we will focus, for clarity, on time-domain results for even spacetime dimensions $d$ (odd $\hd$), where (\ref{S eff leading}) reduces to
\be
S_{eff}^{LO,even} = (-)^{\frac{\hd+1}{2}}\frac{\hd (\hd\!+\!3) (2\!+\!\hd)}{2 \, \hd!!\,(\hd+4)!!}
			 \, \ourG \, \hat{Q}_S^{L_2} \d_t^{\hd+4}Q_S^{L_2}~.
\ee
The LO RR force is given by
\footnote{
Notice a factor of $\frac{1}{\ell!}=\half$ from switching from multi-index notation $L_2$ to usual indices $ij$; see Sec. \ref{Conventions and nomenclature}.
}
\be
F^{i,LO,even}_{SF} = \frac{\delta S_{eff}^{LO,even}}{\delta \hat{x}^i (t)}
= (-)^{\frac{\hd+1}{2}} \, \ourG \, \frac{\hd (\hd\!+\!3) (2\!+\!\hd)}{4 \, \hd!!\,(\hd+4)!!}
	 \frac{\delta \hat{Q}_S^{ij}}{\delta \hat{x}^i (t)} \d_t^{\hd+4}Q_S^{ij}~,
\label{RR force LO}
\ee
and it can be represented by a potential which generalizes the Burke-Thorne potential \cite{BurkeThorne} to any even dimension,
\be
V^{even}_{BT} = (-)^{\frac{\hd+1}{2}} \, \ourG \, m \, x^i x^j \, \frac{\hd (\hd\!+\!3) (2\!+\!\hd)}{4 \, \hd!!\,(\hd+4)!!} \, \d_t^{\hd+4}Q_S^{ij}~.
\label{gen BT}
\ee
The LO dissipated energy is given by
\be
\frac{dE^{LO,even}}{d\w}=\w^{\hd+5} \frac{\hd (\hd\!+\!3) (2\!+\!\hd)}{4 \, \hd!!\,(\hd+4)!!}
			 \, \ourG \left|Q_S^{ij}(\w)\right|^2~,
\label{power w}
\ee
which in the time domain gives the dissipated power as
\be
<P^{LO,even}_{rad}>= \frac{\hd (\hd\!+\!3) (2\!+\!\hd)}{4 \, \hd!!\,(\hd+4)!!}
			 \, \ourG \left<\d_t^{\frac{\hd+5}{2}}Q_S^{ij}(t)\right>^2~,
\label{power t}
\ee
using $Q^S_{ij}(t)$ to LO as given in (\ref{scalar quadrupole leading}).
We note that for $d=4$, (\ref{RR force LO}) and (\ref{gen BT}) reproduce precisely the well-known results for the Burke-Thorne potential and RR force, and that for any $d$, (\ref{power w}) reproduces precisely the generalized Quadrupole formula of Cardoso et al. (\ref{Cardoso even dimension quadrupole}) \cite{Cardoso:2008gn}
\footnote{
In order to revert from $\ourG$ to $G$, see Sec. \ref{Conventions and nomenclature}.
}; (\ref{power t}) matches Einstein's Quadrupole formula in $d=4$ \cite{GWEinstein}, and generalizes it to any even dimension.

Eq. (\ref{S eff leading}) gives the leading order RR effective action (and allows the other quantities to be equally derived from it) in any dimension.
Below we will discuss the higher order contributions; we compute the +1PN correction, and give an outline of higher order contributions.
Although in this section we focus on even $d$, our frequency domain results hold for odd $d$ \footnote{
One may also formally discuss the RR effective action in non-integer $d$, see \cite{PaperIII}.
} as well.
The main difference in odd dimensions appears when Fourier transforming back to the time domain, where due to non-analytic features (branch cuts) of the effective action as a function of (complex) frequency, the transformation introduces non-local (in time) tail terms in the effective action (see Sec. II.C \& IV of \cite{PaperIII} for a thorough discussion and section \ref{Summary} for time-domain results).

\subsection{Next to Leading order (+1PN)}
\label{next to leading}
At the next to leading post-Newtonian order (NLO, which is +1PN), four effects must be considered; we shall label them E3, E2$\delta^1$, E2$\hat{\delta}^1$ and M2, as we now detail.
The scalar quadrupole is supplemented by the scalar octupole ($s\!=\!0,\,\ell\!=\!3$, a.k.a E3), found exactly as in (\ref{S eff leading}), but with $\ell=3$.
In addition, the scalar quadrupole itself becomes more complicated, as it must include +1PN corrections (specialized to the $N$-body problem),
\be
Q^{L_2}_S \to Q^{L_2}_S + \delta^1 Q^{L_2}_S
	= \sum^n_{A=1} m_A\, x_A^{L_2} + \delta^1 Q^{L_2}_S,
\ee
where $\delta^1 Q^{L_2}_S$ includes five possible corrections: the +1PN non-linear corrections to the gravitational mass ($\delta^1 Q_S^{L_2}NL1$), the contribution of the gravitational source current ($\delta^1 Q^{L_2}_S J$), the contribution of a retardation effect from expanding the Bessel function $\tldja(\w r)$ to subleading order ($\delta^1 Q^{L_2}_Sb$), a term with double time derivatives of $\rho_\phi$ ($\delta^1 Q^S_{L_2}\d^2$), and a term with double time derivatives from the derivative of the Bessel function ($\delta^1 Q^{L_2}_S\d b$). Altogether for any $\ell,\hd$ we find the first correction to be
\bea
\label{+1PN corrections source gen ell}
\delta^1 Q^{L}_S
	&=& \delta^1 Q_S^{L}NL1 + \delta^1 Q_S^{L}J + \delta^1 Q_S^{L}b
		+ \delta^1 Q_S^{L}\d^2 + \delta^1 Q_S^{L}\d b \\
	&=& \!\!\! \sum^n_{A=1} \!\! \[ m x^{L} \!\! \( \!\!\!
		\( \frac{\hd\!+\!2}{2\, \hd} v^2
		  -\!\! \sum_{B\neq A} \! \frac{\ourG\, m_B}{\| \vec{x} \!-\! \vec{x}_B\|^\hd}
		\)
		\!\!+ \!\frac{2 (\hd\!+\!1)}{\hd (\ell\!+\!\hd)} i\w \! \( \vec{x} \! \cdot \! \vec{v} \)
		- \frac{\(\! \hd^2\!+\!(\ell\!+\!4)\hd\!+\!4 \)\! \w^2 r^2} {2\hd (\ell\!+\!\hd) (2\ell\!+\!\hd\!+\!2)}
		\) \!\! \]_{\!\!A} \!\! ,  \nonumber
\eea
and specifically for $\ell=2$ (the correction relevant to +1PN order), 
\VERBOSE{	
\be
\delta^1 Q_S^{L_2 \w}
	= \!\! \sum^n_{A=1} \! \[ m x^{L_2} \! \( \!\!
		\( \frac{\hd\!+\!2}{2\, \hd} v^2
		  -\!\! \sum_{B\neq A} \! \frac{\ourG\, m_B}{\| \vec{x} \!-\! \vec{x}_B\|^\hd}
		\)
		\!\!+ \!\frac{2 (\hd\!+\!1)}{\hd (\hd\!+\!2)} i\w \( \vec{x} \! \cdot \! \vec{v} \)
		- \frac{\(\! \hd^2\!+\!6\hd\!+\!4 \!\)\! \w^2 r^2} {2\hd (\hd\!+\!2) (\hd\!+\!6)}
		\) \]_A \!\!.
\label{+1PN corrections source}
\ee
}	
This exactly coincides with (3.109, 3.110) of \cite{PaperI} for $d=4$, as obtained in \cite{Blanchet:1989cu} and reproduced in \cite{GoldbergerRoss} within the EFT approach.
At +1PN, The correction $\delta^1 Q^S_{L_2}$ itself can appear in either the source vertex $Q^S$ (E2$\delta^1$) or in the hatted vertex $\hat{Q}^S$ (E2$\hat{\delta}^1$).

Finally, the +1PN action includes also the leading order vector quadrupole ($s\!=\!1,\,\ell\!=\!2$, a.k.a M2), which includes different $Q^{L_2\w}_{\Valph} \hat{Q}^{L_2\w}_{\Valph}$ terms.
We sum over the $\aleph$ indices to define the bi-vector multipoles
\be
Q^{L\w}_{M} =\frac{1}{\ell\!+\!\hd\!+\!1} \!\! \int\!\!d^Dx
				\[ \vec{r} \wedge
					\( \frac{1}{r^{\ell\!+\!\hd}} \( r^{\ell\!+\!\hd\!+\!1} \tldja(\w r) \)' \!\! \vec{J}
					  + \tldja(\w r) \vec{r}\!\cdot\!\!\dot{\tensor{\Sigma}}
					\)
				\] \!^{(k_\ell} x^{L-1)} .
\label{QM w}
\ee
We accordingly define $\hat{Q}^L_{M}$ , and replace $C\elld^V$ by 
\be
C\elld^M = C\elld^V * \frac{D_\ell (\hd+1,1)}{D_\ell (\hd+1,0)}
\VERBOSE{	
		= C\elld^V * \frac{\ell\hd(\ell+\hd)}{(\ell+1)(\ell+\hd-1)}
		= \Mld \frac{ \ell^2 \, \hd \, (\hd+1) (\ell+\hd+1)}{2 \, (\ell-1)(\ell+1)(\ell+\hd-1)}
}	
~,
\ee
which accounts for the summation over the different $\aleph$ combinations using (\ref{Degeneracies}), following \cite{PaperIII}.
At +1PN order, for the vector contributions we can set $\tldja(\w r)\!=\!1$ and $\tensor{\Sigma}\!=\!0$; the other terms only enter at +2PN and onwards.

With these definitions, the four contributions combine to produce the radiation-reaction effective action to NLO, given by
\bea
S_{eff,\w}^{NLO} &=& -i \w^{\hd+4} \, \ourG \, C^S_{2,\hd} \,
					\( Q^{L_2\w}_S \hat{Q}^{L_2\w \, *}_{S}
					+ \delta^1Q^{L_2\w}_S \hat{Q}^{L_2\w \, *}_{S}
					+ Q^{L_2\w}_S \delta^1\hat{Q}^{L_2\w\,*}_{S} \)	\nonumber\\
&&	-i \w^{\hd+6} \, \ourG \, C^S_{3,\hd} \, Q^S_{L_3\w} \hat{Q}^{S*}_{L_3\w}
	-i \w^{\hd+4} \, \ourG \, C^M_{2,\hd} \, Q^{L_2\w}_{M} \hat{Q}^{L_2\w \, *}_{M} \, + c.c.~~~
\label{S eff next to leading}
\eea

\subsection{Beyond +1PN}
\label{beyond 1PN}
Our method allows identification and enumeration of higher PN order corrections as well. While we do not calculate them explicitly here, it is convenient to see how they too fall under the general prescription.
Effects such as higher-order multipoles ($\ell$), higher-spin ($s$) sectors, retardation effects or non-linearities in the source terms all introduce integer PN order corrections, and will thus only enter at +2PN.
Radiation-zone non-linear corrections and the effects of spin (intrinsic angular momentum) enter at different PN orders depending on the dimension, as we describe below.

\subsubsection*{+2PN}
\label{+2PN}
At +2PN, our formalism captures easily all the relevant near-zone contributions.
The tensor field first enters at this order via the tensor quadrupole ($s\!=\!2,\,\ell\!=\!2$, a.k.a T2).
The vector sector now also includes the vector octupole ($s\!=\!1,\,\ell\!=\!3$, a.k.a M3), as well as +1PN corrections to the vector quadrupole (M2$\delta^1$) and (M2$\hat{\delta}^1$), which arise from the contributions of the gravitational stress ($\delta^1 Q^V_{L_2}\Sigma$), the contribution of two time derivatives of the gravitational current ($\delta^1 Q^V_{L_2}\d^2$), a retardation effect ($\delta^1 Q^V_{L_2}b$), and a derivative of the Bessel function ($\delta^1 Q^V_{L_2}\d b$).
The scalar sector now goes up to the scalar hexadecapole ($s\!=\!0,\,\ell\!=\!4$, a.k.a E4), while corrections to the lower scalar multipoles include five analog corrections to the octupole ($\delta^1 Q^S_{L_3}\d^2$NL1, $\delta^1 Q^S_{L_3}$J, $\delta^1 Q^S_{L_3}\d^2$, $\delta^1 Q^S_{L_3}b$, $\delta^1 Q^S_{L_3}\d b$) and several new corrections to the quadrupole.
For example, the contribution of $T2$ is given by
\be
S^{T2}_{eff,\w} = -i \w^{\hd+4}\, \ourG \, C^T_{2,\hd} \, Q^{L_2 \w}_{\Talbt} \, \hat{Q}^{L_2 \w *}_{\Talbt}.
\ee

\subsubsection*{Radiation zone corrections}
\label{rad zone corrections}
\begin{center}
\begin{figure*}[hbp]
        \begin{center}
            \includegraphics[width=4cm,height=3.5cm]{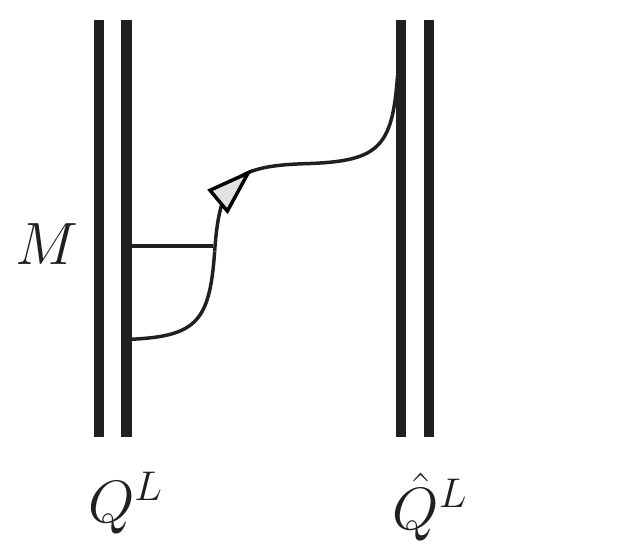} \\
        \end{center}
\caption{Correction to the radiation reaction due to non-linear interaction in the radiation-zone}
\label{radiation non linear}
\end{figure*}
\end{center}

We have thus far taken the propagator to be linear in the radiation zone (\ref{Master Propagator}).
However, there are non-linear corrections to the propagator, which may be included through systematic inclusion of non-linear terms in the action (\ref{S_hom}); see also discussion in section \ref{action_for_pert}.
The first of these contributions is represented by fig. \ref{radiation non linear}, and is interpreted as scattering of the outgoing waves off the background curvature generated by the entire system's total mass $M$ (labeling the vertex in fig. \ref{radiation non linear}).
Its value includes a factor of the gravitational potential $\ourG M / \lambda^\hd$, where $\lambda$ is the typical wavelength for radiation.
As $\lambda\sim\w^{-1}$, the value of such contributions is suppressed by at least $(\w r)^{\hd} \ourG M / r^{\hd} \sim v^{\hd} \ourG M / r^\hd$ relative to the leading order.
In PN terms, this is equivalent to $+(1\!+\!\frac{\hd}{2})$PN order.
In $d=4$, it implies a +1.5PN contribution; it is suppressed even further in higher dimensions.
For further discussion and calculations for the $4d$ case, see \cite{PaperI, Blanchet:1987wq, Blanchet:1993ng, GoldbergerRoss, FoffaSturani4PNa}.
In the high $d$ limit, linearized gravity suffices.

\subsubsection*{Spin Effects}
Depending on the system's parameters and nature of its constituents, spin effects may also need to be considered.
This adds complexity and introduces new interesting effects which are worth exploring also in higher dimensions. 
The important scales determining the PN order - in addition to the typical orbital separation $R$ and typical orbital velocity $v$ - are the typical size of the spinning body $r_0$, its moment of inertia $I\sim m r_0^2$, and its typical angular frequency $\w_s$; together these give the spin
\be
S \sim I\,\w \sim m\, r_0^2 \, \w_s ~.
\label{spin definition}
\ee
Spins can couple to the orbital angular momentum
\be
L \sim m v R~,
\label{orbital angular momentum definition}
\ee
or to each other.
Couplings of the first type (spin-orbit, S-O) appear in the action (at leading order and up to dimensionless factors composed of the masses) as
\be
S_{SO} = \#_{SO} \cdot \int \!\! dt \frac{G}{r^{\hd+2}} S \cdot L
	\sim \int \!\! dt \frac{G m^2}{R^{\hd}} v  \frac{r_0^2 \, \w_s}{R}~,
\label{SO action}
\ee
while couplings of the second type (spin-spin, S-S) appear (again at LO and up to the masses) as
\be
S_{SS} = \#_{SS} \cdot \int \!\! dt \frac{G}{r^{\hd+2}} S \cdot S
	\sim \int \!\! dt \frac{G m^2}{R^{\hd}} \frac{r_0^4 \, \w_s^2}{R^2} ~,
\label{SS action}
\ee
where in the above equations we omit the indices of spin and angular momentum tensors since we are interested only in orders of magnitude.
In terms of the post-Newtonian parameter $v^2$, we recall that $\frac{G m^2}{R^{\hd}}$ is the leading order, and that
\be
\frac{r_0}{R} \gtrsim v^{2 / \hd}~,
\label{pN parameter}
\ee
where similarity occurs in the case of highly compact objects such as black holes (see \cite{MyersPerry}). 
We focus on two interesting cases: co-rotation ($\w_s \sim \w \sim \frac{v}{R}$) and maximal spin ($\w_s\sim \frac{v_s}{r_0}\sim\frac{1}{r_0}$).
In the case of co-rotation, we find that S-O effects enter with a suppression of $v^2\(\frac{r_0}{R}\)^2$ corresponding to PN order $1\!+\!2/\hd$, while S-S effects enter with a suppression of $v^2\(\frac{r_0}{R}\)^4$, which means PN order $1\!+\!4/\hd$.
For maximal spin, S-O coupling incurs a suppression of $v\frac{r_0}{R}$ ($0.5\!+\!1/\hd$ PN) and S-S a suppression of $\(\frac{r_0}{R}\)^2$ ($2/\hd$ PN).
Thus for co-rotation, spin effects only enter beyond +1PN in any dimension.
For maximal rotation, while in $4d$ S-O enters at +1.5PN and S-S at +2PN (compare \cite{BlanchetRev, PortoSpin2005, PortoRothstein2006, BarkerO'Connel1, BarkerO'Connel2, Kidder:1992fr, Kidder:1995zr, Owen:1997ku, Tagoshi:2000zg, Steinhoff:2007mb, Levi:2008nh, Barausse:2009aa}), as the dimension grows both effects become more and more important, with both effects at +1PN for $d=5$, and both entering before +1PN at $d>5$.
Also for $d>5$, spin-spin interactions become more dominant than spin-orbit.

\section{Summary of results}
\label{Summary}
In general spacetime dimension $d$ the radiative field decouples, at the linear level, into three sectors - scalar, vector, and tensor - with respect to the $\Omd$ sphere.
The radiation is generated by the corresponding multipole moments of the source.
The linearized-GR multipoles are given in the time domain by
\bea
Q^L_S &=& \frac{(\hd\!+\!1)}{\hd (\ell\!+\!\hd) (\ell\!+\!\hd\!+\!1)} \! \int \!\! d^D x \, x^{STF}_L \!
\[	\hd \( \frac{(\ell\!+\!\hd)(\ell\!+\!\hd\!+\!1)}{\hd+1} + r\d_r + \frac{r^2}{\hd}\d_t^2\) \!
		 \tldja(i r \d_t) T^{tt}
	\right. \nonumber \\ &&~~~~~~~~~~~~~~~~~~~~~~~~~~~~~~~~~~~~~   \left.
	-2 \, x_a \( \ell\!+\!\hd\!+\!1+r\d_r \) \tldja(i r \d_t) \d_t T^{ta}
	\right. \nonumber \\ &&~~~~~~~~~~~~~~~~~~~~~~~~~~~~~~~~~~~~~   \left.
	+\( \frac{(\ell\!+\!\hd)(\ell\!+\!\hd\!+\!1)}{\hd+1} + r\d_r \)
		\tldja(i r \d_t) T^{aa}
	\right. \nonumber \\ &&~~~~~~~~~~~~~~~~~~~~~~~~~~~~~~~~~~~~~   \left.
	+ x_a x_b \, \tldja(i r \d_t) \d_t^2 T^{ab}
\] ~,
\label{QS}	\\
Q^L_{\Valph}&=&\frac{2\eps^{(D)}_{\aleph a b k_\ell} }{\ell\!+\!\hd\!+\!1} \!\! \int\!\!d^Dx\! 
					\[ \( r^{\ell\!+\!\hd\!+\!1} \tldja(i r \d_t) \)'
						\frac{x^{bL-1}}{r^{\ell\!+\!\hd}} T^{ta}
					  \!-\! \tldja(i r \d_t) \d_t T^{ac} x^{bcL-1}
					\],~~~~~~~
\label{QV}	\\
Q^L_{\Talbt}
&=& \frac{\ell(\ell\!-\!1)}{2}
		\eps^{(D)}_{\aleph a b k_{\ell}} \eps^{(D)}_{\beth a' b' k_{\ell'}} 
		\int \!\! d^D x\, x^{bb' \! L-2}  T^{aa'} \tldja(i r \d_t) \,
\label{QT}~,
\eea
where $T^{\mu \nu}(\vec{r},t)$ is the energy momentum tensor.
It is useful to sum over the $\aleph$ coordinates to replace the vector multipoles by
\be
Q^L_{M} =\frac{1}{\ell\!+\!\hd\!+\!1} \!\! \int\!\!d^Dx
				\[ \vec{r} \wedge
					\( \frac{1}{r^{\ell\!+\!\hd}} \( r^{\ell\!+\!\hd\!+\!1} \tldja(i r \d_t) \)' \!\! \vec{J}
					  + \tldja(i r \d_t) \vec{r}\!\cdot\!\!\dot{\tensor{\Sigma}}
					\)
				\] \!^{(k_\ell} x^{L-1)}~.
\label{QM}
\ee
These multipoles share the same form in any spacetime dimension.
The Feynman propagator (\ref{Master Propagator}) introduces $2\ell+\hd$ factors of $\w$ into the RR effective action (\ref{S eff general}).
In even dimensions, these are transformed to the time domain to become $2\ell+\hd$ time derivatives.

For the N-body problem at leading post-Newtonian order, the radiation-reation effective action includes only the leading order of the scalar quadrupole $Q^{ij}_{S} = \sum^n_{A=1} m_A (x_A^{ij})^{TF}$ (\ref{scalar quadrupole leading}), and the action reproduces the generalized Burke-Thorne potential, as well as eq. (\ref{Cardoso even dimension quadrupole}), as described in Sec. \ref{LO}.

At +1PN order, contributions arise from the scalar (mass) octupole,
\be
Q^{L_3}_{S} = Q^{ijk}_{S} = \sum^n_{A=1} m_A \[ x^i x^j x^k
									-\frac{1}{D+2}\(\delta^{ij}x^k+\delta^{ik}x^j+\delta^{jk}x^i\)\!x^2
									\]_A~,
\ee
from the vector (current) quadrupole,
\be
Q^{L_2}_{M} = Q^{ij}_{M} = 2 \sum^n_{A=1} \[ m \( \vec{r} \wedge \vec{J} \) \!^{(i} x^{j)} \]_A ~,
\label{QM quadrupole}
\ee
and from corrections to the scalar quadrupole,
\bea
\delta^1 Q_S^{L_2} &=& \!\! \sum^n_{A=1} \! m_A \! \[ \!
		\( \frac{\hd\!+\!2}{2\, \hd} v_A^2
		 - \sum_{B\neq A} \! \frac{\ourG\, m_B}{\| \vec{x}_A \!-\! \vec{x}_B\|^\hd}
		\) \! \! x_A^{L_2}
		- \frac{2 (\hd\!+\!1)}{\hd (\hd\!+\!2)} \d_t \( \vec{x}_A \! \cdot \! \vec{v}_A \, x_A^{L_2} \)
\right.\nonumber\\
&& \left. ~~~~~~~~~
		+ \frac{\(\! \hd^2\!+\!6\hd\!+\!4 \!\) } {2\hd (\hd\!+\!2) (\hd\!+\!6)}
			\d_t^2 \( r_A^2 \, x_A^{L_2}\)
		\] ~.
\eea
Taken together we find the radiation-reaction effective action to +1PN, which in the case of even dimension $d$ is
\bea
S_{eff}^{NLO} &=& (-)^{\frac{\hd+1}{2}} \ourG \!\! \int \!\! dt \!
		\[
			\frac{\hd (\hd+2) (\hd+3)}{2 \, \hd!! \, (\hd+4)!!}
				\( \hat{Q}^{L_2}_S \d_t^{d+1} Q^{L_2}_S
				+ \hat{Q}^{L_2}_S \d_t^{d+1} \delta^1Q^{L_2}_S
				+ \delta^1\hat{Q}^{L_2}_S \d_t^{d+1} Q^{L_2}_S
				\)
		\right.
\nonumber\\
&&	~~~~~~~~~~~~~~~~~~
		\left.
			- \frac{\hd (\hd+4) (\hd+3)}{6 \, \hd!! (\hd+6)!!}
				\, \hat{Q}^{L_3}_{S} \d_t^{d+3} Q^{L_3}_S
			+ \frac{ 2 \, \hd \, (\hd+3)}{3 \, \hd!! \, (\hd+4)!!} \,
				\hat{Q}^{L_2}_M \d_t^{d+1} Q^{L_2}_M
		\]
\VERBOSE{	
	\nonumber\\
&=& (-)^{\frac{\hd+1}{2}} \ourG
				\frac{\hd (\hd+2) (\hd+3)}{4 \, \hd!! \, (\hd+4)!!}
					\( \hat{Q}^{ij}_S \d_t^{d+1} Q^S_{ij}
					+ \hat{Q}^{ij}_S \d_t^{d+1} \delta^1Q^{ij}_S
					+ \delta^1\hat{Q}^{ij}_S \d_t^{d+1} Q^{ij}_S \)
\nonumber\\
&&	+ (-)^{\frac{d}{2}} \ourG
		\frac{\hd (\hd+4) (\hd+3)}{36 \, \hd!! (\hd+6)!!}
			\, \hat{Q}^{ijk}_S \d_t^{d+3} Q^{ijk}_S
	+ (-)^{\frac{\hd+1}{2}} \ourG \frac{\hd \, (\hd+3)}{3 \, \hd!! \, (\hd+4)!!} \,
			\hat{Q}^{ij}_M \d_t^{d+1} Q^{ij}_M
	\nonumber\\
}	
\label{S eff next to leading even time}.~~~~~~~
\eea
This is the main result of this paper.
It matches (3.111) of \cite{PaperI} in $d=4$ (upon re-introduction of $1/\ell!$ from the summation convention), and extends it to different dimensions.
The energy dissipated from the system, as well as the gravitational self-force acting on it, can be read off from this equation.

While all the results of this paper, presented in the frequency domain, are valid for any spacetime dimension, for odd dimensions the Fourier transformation to the time domain introduces non-local ``tail" expressions in the action.
This happens due to the appearance of branch cuts in the frequency domain effective action.
For a thorough discussion of this, see eq. (4.7, 4.8, 4.14, 4.15) of \cite{PaperIII}; as an example, the time-domain effective action in odd $d$ is given by
\bea
\hS_{eff}&=&\ourG \! \int\!\!{dt}\!\sum_{L} (-)^{\ell+\frac{\hd+1}{2}}
	 \[
		C\elld^S \, S^{(S)}(t) + C\elld^M \, S^{(M)}(t) + C\elld^T \, S^{(T)}(t)
	 \] ,~~~~
\label{S multipoles odd dimension 1}
\\  \nonumber\\ 
S^{\eps}(t) &=& \hat{Q}^{\eps}_L(t)
\[
	\(\frac{1}{2}H(2\ell+\hd)-H(\ell+\frac{\hd}{2}) \) \d_t^{2\ell+\hd}Q_{\eps}^L(t)  \right. \nonumber\\
&& \left. \left. ~~~~~~~~~~~~~~
	-\int_{-\infty}^{t}\!\!\!\! dt' \( \frac{1}{t-t'} \d_{t'}^{2\ell+\hd}Q_{\eps}^L(t') \) \right|_{regularized}
\] , ~
\label{S multipoles odd dimension 2}
\eea
where $\eps\in\{S,V,T\}$, $H(n)$ is the $n$'th Harmonic Number, and regularization (hence the effective coefficient of the local part) depends on shot-distance details of the system, in particular on the scale (see also \cite{PaperIII, BlanchetGenD, Blanchet:2002av, Goldberger:2007hy}).

\subsection*{Acknowledgments}
This research was supported by the Israel Science Foundation, grant no. 812/11, and it is part of the Einstein Research Project ``Gravitation and High Energy Physics", which is funded by the Einstein Foundation Berlin.
OB was partly supported by an ERC Advanced Grant to T. Piran.
We are grateful to R.~Emparan, A.~Harte and J.~Steinhoff for engaging discussions, and especially thank B.~Kol for support, advice and comments.

\appendix

\VERBOSE{	
\section{Quadratic Action}
\label{app:quadratic action}
From the definitions,
\bea
\Gm_{\al\mu\nu} &=& \frac{1}{2} \( h_{\al\mu,\nu} + h_{\al\nu,\mu} - h_{\mu\nu,\al} \)~~;~~
\Gm^\alpha_{\mu\nu}
	= \frac{g^{\alpha\bt}}{2} \( g_{\mu\bt,\nu} + g_{\nu\bt,\mu} -g_{\mu\nu,\bt} \)
	= g^{\al\bt} \Gm_{\bt\mu\nu}~~~~~~\\
R_{iklm} &=& \frac{1}{2}\[ g_{im,kl} + g_{kl,im} - g_{il,km} - g_{km,il} \]
	+ g_{np} \[ \Gm^n_{kl} \Gm^p_{im} - \Gm^n_{km} \Gm^p_{il} \]	\nonumber\\
	&=& \frac{1}{2}\[ g_{im,kl} + g_{kl,im} - \( l \leftrightarrow m\) \]
	+ g_{np} \[ \Gm^n_{kl} \Gm^p_{im} - \( l \leftrightarrow m\) \]\\
R&=&g^{il} g^{km} R_{iklm}
\eea
We expand up to quadratic order in $h$ (recalling $R_{iklm}$ starts at linear order):
\bea
\int \!\! d^d x \detg R
	&=& \int \!\! d^d x\detg  g^{il} g^{km} R_{iklm}
	= \int \!\! d^d x \( 1+\frac{1}{2}h_{\al\al} \)
		\( \eta^{il} - h^{il}\) \( \eta^{km} - h^{km}\) R_{iklm} \nonumber\\
	&=& \int \!\! d^d x
		\( \eta^{il}\eta^{km}
			+\frac{1}{2}h_{\al\al}\eta^{il}\eta^{km}
			- \eta^{il}h^{km} -\eta^{km}h^{il}
		\)
		R_{iklm}
\eea
Expanding both parts of $\int \!\! d^d x \detg R$ to quadratic order and with integration by parts, we find first:
\bea
\int \!\! d^d x && \!\!\!\!\detg  g^{il} g^{km}\frac{1}{2} \[ g_{im,kl} + g_{kl,im} - \( l \leftrightarrow m\) \] \nonumber\\
&&\simeq
 \int \!\! d^d x
		\( \eta^{il}\eta^{km}
			\!+\! \frac{1}{2}h_{\al\al}\eta^{il}\eta^{km}
			\!-\! \eta^{il}h^{km} \!-\! \eta^{km}h^{il}
		\)
	\frac{1}{2} \[ h_{im,kl} + h_{kl,im} - \( l \leftrightarrow m\) \]~~~~~~~~~\nonumber\\
&&=
\frac{1}{2} \int \!\! d^d x
		\[
			+\frac{1}{2}h_{\al\al}\eta^{il}\eta^{km} h_{im,kl}
			- \eta^{il}h^{km}h_{im,kl} - \eta^{km}h^{il}h_{im,kl}
		\right. \nonumber\\ && \left.~~~~~~~~~~~~~~
			+ \frac{1}{2}h_{\al\al}\eta^{il}\eta^{km}  h_{kl,im}
			-\eta^{il}h^{km} h_{kl,im} - \eta^{km}h^{il} h_{kl,im}
		\right. \nonumber\\ && \left.~~~~~~~~~~~~~~
			-\frac{1}{2}h_{\al\al}\eta^{il}\eta^{km} h_{il,km}
			+ \eta^{il}h^{km}h_{il,km} + \eta^{km}h^{il}h_{il,km}
		\right. \nonumber\\ && \left.~~~~~~~~~~~~~~
			- \frac{1}{2}h_{\al\al}\eta^{il}\eta^{km}  h_{km,il}
			+\eta^{il}h^{km} h_{km,il} + \eta^{km}h^{il} h_{km,il}
		\]~~~~~~~~~
\nonumber\\
&&=\frac{1}{2} \int \!\! d^d x
		\[
			+h_{\al\al,k} h_{ii,k}
			+ 4h_{mk,k}h_{mi,i}
			-3h_{\al\al,k} h_{ki,i}
			-2h_{il,k}h_{il,k}
		\]~~~~~~~~~
\eea
and second:
\bea
\int \!\! d^d x && \!\!\!\!\detg  g^{il} g^{km} g_{np} \[ \Gm^n_{kl} \Gm^p_{im} - \( l \leftrightarrow m\) \] \simeq
	 \int \!\! d^d x \, \eta^{il}\eta^{km} g_{np}
		\[ \Gm^n_{kl} \Gm^p_{im} - \( l \leftrightarrow m\) \]~~~~~~~~~
	\nonumber\\
&&\simeq
	 \int \!\! d^d x \, \frac{1}{4}\eta^{il}\eta^{km}
		\[
	 \( h_{k\bt,l} + h_{l\bt,k} -h_{kl,\bt} \) \( h_{i\bt,m} + h_{m\bt,i} -h_{im,\bt} \)
		 - \( l \leftrightarrow m\) \]~~
	\nonumber\\
&&=
	 \int \!\! d^d x \, \frac{1}{4}\eta^{il}\eta^{km}
		\[
		+h_{k\bt,l}h_{i\bt,m} +h_{k\bt,l}h_{m\bt,i} -h_{k\bt,l}h_{im,\bt}
			\right. \nonumber\\ && \left.
		~~~~~~~~~~~~~~~~~~~~~~ +h_{l\bt,k}h_{i\bt,m} + h_{l\bt,k}h_{m\bt,i} -h_{l\bt,k}h_{im,\bt}
			\right. \nonumber\\ && \left.
		~~~~~~~~~~~~~~~~~~~~~~-h_{kl,\bt}h_{i\bt,m} -h_{kl,\bt}h_{m\bt,i} +h_{kl,\bt}h_{im,\bt}
			\right. \nonumber\\ && \left.
		~~~~~~~~~~~~~~~~~~~~~~-h_{k\bt,m}h_{i\bt,l} -h_{k\bt,m}h_{l\bt,i} +h_{k\bt,m}h_{il,\bt}
			\right. \nonumber\\ && \left.
		~~~~~~~~~~~~~~~~~~~~~~ -h_{m\bt,k}h_{i\bt,l} - h_{m\bt,k}h_{l\bt,i} +h_{m\bt,k}h_{il,\bt}
			\right. \nonumber\\ && \left.
		~~~~~~~~~~~~~~~~~~~~~~+h_{km,\bt}h_{i\bt,l} +h_{km,\bt}h_{l\bt,i} -h_{km,\bt}h_{il,\bt}
	 \]~~
	\nonumber\\
&&=
	 \int \!\! d^d x \, \frac{1}{4}
		\[+3h_{k\bt,i}h_{k\bt,i}+4h_{ii,\bt}h_{k\bt,k}-h_{kk,\bt}h_{ii,\bt}-6h_{i\bt,\bt}h_{ik,k}
	 \]~~
\eea
Together we have
\be
\int \!\! d^d x \detg  R
\simeq  \int \!\! d^d x
	\[-\frac{1}{4}h_{k\bt,i}h_{k\bt,i}
	  -\frac{1}{2}h_{ii,\bt}h_{k\bt,k}
	   +\frac{1}{4}h_{kk,\bt}h_{ii,\bt}
	  +\frac{1}{2}h_{i\bt,\bt}h_{ik,k}
	 \]
\ee
Alternatively,
\bea
2 \gm^\lambda_{\,\,\mu\nu}\!\!\!\!\!\!\!\!\! &&\gm_{\,\,\,\,\lambda}^{\mu\nu} + \frac{1}{2(\hd+1)} \nabla_\mu\hbar\nabla^\mu\hbar	\nonumber\\
	&=& \frac{1}{2}
	\( \d_\mu \hbar_{\lambda\nu} + \d_\nu \hbar_{\lambda\mu} -\d_\lambda \hbar_{\mu\nu} \)
	\( \d_\nu \hbar_{\lambda\mu} + \d_\lambda \hbar_{\mu\nu} - \d_\mu \hbar_{\lambda\nu}\)
	+ \frac{(\hd+1)^2}{8(\hd+1)} \d_\nu h_{\mu\mu}  \d_\nu h_{\alpha\alpha}	~~~~~~~~~~~~ \nonumber\\
	&=& +\frac{\hd+1}{8} \d_\nu h_{\mu\mu}  \d_\nu h_{\alpha\alpha}
	+\frac{1}{2}
	\[2\d_\lambda \hbar_{\mu\nu}\d_\mu \hbar_{\lambda\nu}  -(\d_\mu \hbar_{\lambda\nu})^2
	\]~~~~~~~~~~~~
\nonumber\\
	&=& \frac{\hd+1}{8} \d_\nu h_{\mu\mu}  \d_\nu h_{\alpha\alpha}
	+\d_\lambda h_{\mu\nu}\d_\mu h_{\lambda\nu}
	+\frac{1}{4} \d_\nu h_{\al\al} \d_\nu h_{\mu\mu}
	-\d_\mu h_{\al\al} \d_\nu h_{\mu\nu} \!
	-\frac{1}{2}(\d_\mu h_{\lambda\nu})^2
	+\frac{4-d}{8}\d_\gamma h_{\al\al} \d_\gamma h_{\beta\beta}
\nonumber\\
	&=& \frac{1}{2} \d_\nu h_{\mu\mu}  \d_\nu h_{\alpha\alpha}
	+ \d_\mu h_{\mu\nu} \d_\lambda h_{\lambda\nu}
	-\d_\nu h_{\al\al} \d_\mu h_{\mu\nu} \!
	-\frac{1}{2}(\d_\mu h_{\lambda\nu})^2
\eea
which is exactly twice the integrand $\int \!\! d^d x \detg R$.
}	

\section{Properties of spherical derivatives}
\label{app:spherical harmonics}
The generalized spherical metric for coordinates $\{t,r,\Om_1,\Om_2,\ldots,\Om_{\hd+1}\}$ is
\be
g_{\mu\nu}=diag\{-1,1,r^2\Pi_1,r^2\Pi_2,\ldots,r^2\Pi_{\hd+1}\},
	~~~~~\Pi_i=\prod^{i-1}_{j=1}\sin^2\Om_j,
\label{metric in spherical variables}
\ee
from which we find all the Christoffel symbols
\bea
\Gm^r_{\Om_a\Om_b}&=&-r g_{\Om_a\Om_b},~~~~
\Gm^{\Om_a}_{r\Om_b}=\Gm^{\Om_b}_{r\Om_a}
	=\frac{1}{r}g^{\Om_a}_{\Om_b}=\frac{1}{r}\delta^{\Om_a}_{\Om_b},
~~~~ \Gm^t_{\al\bt}=\Gm^\al_{t\bt}=0,
\nonumber\\
\Gm^{\Om_a}_{\Om_b\Om_c}&=&
	\delta^{\Om_a}_{\Om_b}\theta(a-c)cot\Om_c
	+\delta^{\Om_a}_{\Om_c}\theta(a-b)cot\Om_b
	-\delta^{\Om_b}_{\Om_c}\theta(b-a)\frac{\Pi_b}{\Pi_a}\cot\Om_a,
\label{spherical christoffels}
\eea
where $\theta(a-b)\!=\!1$ for $a\!>\!b$, otherwise $0$.
The Riemann tensor on the sphere is
\be
R_{\Om_a\Om_b\Om_c\Om_d}=
	g_{\Om_a\Om_c}g_{\Om_b\Om_d} - g_{\Om_a\Om_d}g_{\Om_b\Om_c} \,.
\ee
We shall also make use of the relation
\be
g^{\Om\Om'}g_{\Om\Om'}=\hd+1,
\ee
and note that
\be
g_{\Om\Om',\Om}=0.
\ee
The volume element is
\be
\sqrt{-g}\, d^d x = r^{\hd+1}\prod_{i=1}^\hd \(\sin\Om_i\)^{\hd+1-i} dtdrd\Om_1\!\cdots\!\Om_{\hd+1} \triangleq dt\, r^{\hd+1}dr \, \dOmd.
\ee
Using the Riemann tensor we find the commutation relations for a tensor on the sphere:
\bea
\[ D_{\Om_a} , D_{\Om_b} \] V_{\Om_c} = R^{\Om_d}_{~\Om_c\Om_b\Om_a}\!\!\!&&\!\!\!\! V_{\Om_d}
	= g_{\Om_c [ \Om_a} V_{\Om_b ]}
\VERBOSE{	
\\
\[ D_{\Om''} , D_{\Om} \] x_{\Om'} = g_{\Om' [ \Om''} x_{\Om ]}
~~&,&~~
\[ D_{\Om''} , D_{\Om} \] D_{\Om'} x = g_{\Om' [ \Om''} D_{\Om ]} x \\
g^{\Om'\Om''}\[ D_{\Om'} , D_{\Om} \] x_{\Om''} = \hd \, x_{\Om}
~~&,&~~
g^{\Om'\Om''}\[ D_{\Om'} , D_{\Om} \] D_{\Om''} x = \hd \, D_{\Om} x
\eea
And also
\bea
\[ D_{\Om''} , D_{(\Om} \] x_{\Om')} &=& g_{\Om' [ \Om''} x_{\Om ]} + g_{\Om [ \Om''} x_{\Om' ]}
	= g_{\Om'\Om''} x_{\Om} + g_{\Om\Om''} x_{\Om'} -2g_{\Om\Om'} x_{\Om''} ~~~~~~~ \\
D^{\Om''} \[ D_{\Om''} , D_{(\Om} \] x_{\Om')} &=&
	D_{\Om'}x_{\Om} + D_{\Om}x_{\Om'} -2g_{\Om\Om'} D^{\Om''} x_{\Om''} = D_{(\Om}x_{\Om')}
}	
\eea
The commutators for tensors give
\bea
\[ D_{\Om_a} , D_{\Om_b} \] T_{\Om_c\Om_d}
\VERBOSE{	
&=& R^{\Om_e}_{~\Om_c\Om_b\Om_a}T_{\Om_e\Om_d} + R^{\Om_e}_{~\Om_d\Om_b\Om_a}T_{\Om_c\Om_e}
\nonumber\\
}	
&=& g_{\Om_a\Om_c}T_{\Om_b\Om_d} -g_{\Om_b\Om_c}T_{\Om_a\Om_d}
+g_{\Om_a\Om_d}T_{\Om_b\Om_c} -g_{\Om_b\Om_d}T_{\Om_a\Om_c} ~,~~~~~~\\
\[ D_{\Om''} , D_{ Q} \] \( D_{\Om''} V_P \)
\VERBOSE{	
	&=& g_{\Om''\Om''}D_{Q}V_{P} -g_{Q\Om''}D_{\Om''}V_{P}
	+g_{\Om''P}D_{Q}V_{\Om''} -g_{QP}D_{\Om''}V_{\Om''}
\nonumber\\
	&=& (\hd+1)D_{Q}V_{P} -D_{Q}V_{P}
	+D_{Q}V_{P} -g_{QP}D_{\Om''}V_{\Om''}
\nonumber\\
}	
	&=& (\hd+1)D_{Q}V_{P} -g_{QP}D_{\Om''}V_{\Om''}
\VERBOSE{	
\\
\[ D_{\Om''} , D_{ Q} \] \( D_{\Om''} x_P \) &=&(\hd+1)D_{Q}x_{P}\\
\[ D_{\Om''} , D_{ Q} \] \( D_{\Om''} D_P x \) &=& \( (\hd+1)D_{Q} D_{P} +c_s g_{QP} \)x \\
\[ D_{\Om''} , D_{ (\Om} \] x_{\Om');\Om''}
\VERBOSE{	
&=&
	\[ D_{\Om''} , D_{ \Om} \] \( D_{\Om''} x_{\Om'} \) + \[ D_{\Om''} , D_{ {\Om'}} \] \( D_{\Om''} x_{\Om} \)=\nonumber\\
}	
	&=& (\hd+1)D_{\Om}x_{\Om'} + (\hd+1)D_{\Om'}x_{\Om}=(\hd+1)D_{(\Om}x_{\Om')}
}	
~.
\eea

\VERBOSE{	

The decomposition to spherical harmonics is:
\bea
h_{tt} &=& \int\!\!\frac{d\w}{2\pi}e^{-i\w t} \!\!\!\!\!\! \sum_{\ell,m_1 \ldots m_n}\!\!\!\!\!\!  h^S_{tt,\ell,m_1\ldots m_n} Y^S_{\ell,m_1\ldots m_n} \nonumber\\
h_{tr} &=& \sumint e^{-i\w t} h^S_{tr} Y^S~~~~~~~~~~~~~~~~~,~~
h_{rr} = \sumint e^{-i\w t} h^S_{rr} Y^S \nonumber\\
h_{ti} &=& \sumint e^{-i\w t} \( h^S_t Y_i^S + h^V_t Y_i^V \) ~~,~~
h_{ri} = \sumint e^{-i\w t} \( h^S_r Y_i^S + h^V_r Y_i^V \) \nonumber\\
h_{ij} &=& \sumint e^{-i\w t} \( h^S Y_{ij}^S + \tilde{h}^S \tilde{Y}_{ij}^S + h^V Y_{ij}^V + h^T Y_{ij}^T \)
\eea
They satisfy the following derivative properties
\bea
\d_t h &=& -i\w h ~~,~~
	\d_r h = h' ~~,~~
	\d_\Om h = 0 ~,~~ \nonumber\\
	\Lapd Y^S &=& -c_s Y^S ~,~~ \nonumber\\
Y^S_\Om &=& \d_\Om Y^S ~~,~~
	\d_\Om Y^S_\Om = -c_s Y^S  ~~,~~
	\Lapd Y^S_\Om = -(c_s-\hd) Y^S_\Om ~,~~ \nonumber\\
Y^S_{\Om\Om'}&=&g_{\Om\Om'}Y^S ~~,~~
	g_{\Om\Om'} Y^S_{\Om\Om'} = (\hd\!+\!1) Y^S ~,~
	\d_\Om Y^S_{\Om\Om'} = Y^S_{\Om'} ~~,~~
	\Lapd Y^S_{\Om\Om'} = -c_s Y^S_{\Om\Om'}~,~~\nonumber\\
\tilde{Y}^S_{\Om\Om'}&=&\(D_\Om D_{\Om'} +\frac{c_s}{\hd\!+\!1} g_{\Om\Om'} \) Y^S ~,~
	g_{\Om\Om'} \tilde{Y}^S_{\Om\Om'} = 0 ~,~
	\d_\Om \tilde{Y}^S_{\Om\Om'} = -\frac{\hd \, \hc_s}{\hd\!+\!1} Y^S_{\Om'} ~,~~ \nonumber\\
\Lapd \tilde{Y}^S_{\Om\Om'} &=& -(c_s-2(\hd\!+\!1)) \tilde{Y}^S_{\Om\Om'} ~,~~ \nonumber\\
\d_\Om Y^V_\Om &=& 0 ~~,~~
	\Lapd Y^V_\Om = -(c_s - 1) Y^V_\Om  ~~,~~ \nonumber\\
Y^V_{\Om\Om'}&=&\frac{1}{2}\( D_\Om Y^V_{\Om'} + D_{\Om'} Y^V_\Om \) ~,~
	\d_\Om Y^V_{\Om\Om'} = -\frac{\hc_s}{2} Y^V_{\Om'} ~,~
	\Lapd Y^V_{\Om\Om'} = -(c_s - d) Y^V_{\Om\Om'}	~,~~ \nonumber\\
\d_\Om Y^T_{\Om\Om'} &=& 0 ~~,~~
	\Lapd Y^T_{\Om\Om'} = -(c_s - 2) Y^T_{\Om\Om'}~,~~
\eea
which are derived by calculating commutation relations on the sphere above, as well as these $\Lapd$ eigenvalues
\bea
\Lapd Y^V_{\Om\Om'}
\VERBOSE{	
&=& D_{\Om''} D_{\Om''} \frac{ D_{(\Om} Y^V_{\Om')} }{2}
\VERBOSEE{	
	= \frac{1}{2} \( D_{\Om''} D_{\Om''} D_\Om Y^V_{\Om'} + D_{\Om''} D_{\Om''} D_{\Om'} Y^V_\Om \) \nonumber\\
&=& \frac{1}{2}
	\( D_{\Om''} D_\Om D_{\Om''} Y^V_{\Om'} + D_{\Om''} \[ D_{\Om''} , D_\Om \]  Y^V_{\Om'}
	+ D_{\Om''} D_{\Om'} D_{\Om''} Y^V_\Om + D_{\Om''} \[ D_{\Om''} , D_{\Om'} \] Y^V_\Om \) \nonumber\\
&=& \frac{1}{2}
	\( D_{(\Om} \(Y^V_{\Om'); \Om''\Om''}\) + [ D_{\Om''}, D_{(\Om} ] Y^V_{\Om'); \Om''}
	+ D_{\Om''} \[ D_{\Om''} , D_{(\Om} \]  Y^V_{\Om')} \) \nonumber\\
&=& \frac{1}{2}
	\( -(c_s-1) D_{(\Om} Y^V_{\Om')} +(\hd+1) D_{(\Om} Y^V_{\Om')} +D_{(\Om} Y^V_{\Om')} \)
}	
\nonumber\\
}	
	&=& -\frac{c_s-\hd-3}{2} D_{(\Om} Y^V_{\Om')} = -(c_s-d) Y^V_{\Om\Om'}
\eea
and
\bea
\Lapd \tilde{Y^S}_{\Om\Om'}
\VERBOSE{	
&=& g^{\Om''\Om'''}D_{\Om'''}D_{\Om''} \(D_\Om D_{\Om'} +\frac{c_s}{\hd\!+\!1} g_{\Om\Om'}\) Y^S \nonumber\\
\VERBOSEE{	
&=& g^{\Om''\Om'''}D_{\Om'''}D_{\Om''} D_\Om D_{\Om'} Y^S
	+\frac{c_s}{\hd\!+\!1} g_{\Om\Om'} g^{\Om''\Om'''}D_{\Om'''}D_{\Om''} Y^S \nonumber\\
&=& g^{\Om''\Om'''} D_{\Om'''} D_\Om D_{\Om''} D_{\Om'} Y^S
	+g^{\Om''\Om'''} D_{\Om'''} [ D_{\Om''}, D_\Om]  D_{\Om'} Y^S
	+\frac{c_s}{\hd\!+\!1} g_{\Om\Om'} \Box Y^S 				\nonumber\\
&=& g^{\Om''\Om'''} D_{\Om'''} D_\Om D_{\Om''} D_{\Om'} Y^S
	\!\!+\!g^{\Om''\Om'''} D_{\Om'''} \( g_{\Om''\Om'}D_\Om \!-\!g_{\Om\Om'}D_{\Om''} \) Y^S
	\!\!-\!\frac{c_s^2}{\hd\!+\!1} g_{\Om\Om'} Y^S 					\nonumber\\
&=& g^{\Om''\Om'''} [ D_{\Om'''} , D_\Om ] D_{\Om''} D_{\Om'} Y^S
	+ g^{\Om''\Om'''} D_\Om D_{\Om'''} D_{\Om''} D_{\Om'} Y^S
	\nonumber\\	&&
	+\( D_{\Om'}D_\Om -g_{\Om\Om'} \Box \) Y^S
	-\frac{c_s^2}{\hd\!+\!1} g_{\Om\Om'} Y^S 					\nonumber\\
&=& \( (\hd+1) \, D_{\Om} D_{\Om'} + c_s g_{ \Om\Om'} \) Y^S
	+ g^{\Om''\Om'''} D_\Om D_{\Om'''} D_{\Om'} D_{\Om''} Y^S		\nonumber\\
	&& +\( c_s g_{\Om\Om'} + D_{\Om'}D_\Om \) Y^S
	-\frac{c_s^2}{\hd\!+\!1} g_{\Om\Om'} Y^S \nonumber\\
&=& \( \! (\hd+2) \, D_{\Om} D_{\Om'} \!-\! c_s\frac{c_s-2\hd-2}{\hd\!+\!1}
		g_{\Om\Om'}\! \)\!\! Y^S
	\nonumber\\	&&~~~~~~~
	+ g^{\Om''\Om'''} D_\Om \( [D_{\Om'''},D_{\Om'}] D_{\Om''}
		+ D_{\Om'} D_{\Om'''} D_{\Om''} \)\! Y^S				\nonumber\\
&=& \( (\hd+2) \, D_{\Om} D_{\Om'} - c_s\frac{c_s-2(\hd+1)}{\hd+1} g_{\Om\Om'} \) Y^S
	+\hd\, D_\Om D_{\Om'} Y^S - c_s D_\Om D_{\Om'} Y^S			\nonumber\\
}	
&=& -\( c_s \!-\! 2(\hd\!+\!1) \) \( D_{\Om} D_{\Om'} + \frac{c_s}{\hd+1} g_{\Om\Om'} \) Y^S
}	
	= -\( c_s \!-\! 2(\hd\!+\!1) \) \tilde{Y}^S~~~~~~~
\eea

}	

\section{Spherical fields}
\label{app:Spherical fields}
For a rank-2 tensor $A_{\mu\nu}$ in d dimensions we use the following spherical decomposition ($A$'s $L$, $\w$ indices are suppressed):
\bea
A_{\al\bt}&=&
	\begin{pmatrix}
		  A_{tt} & A_{tr} & A_{t\Om} \\
		  A_{tr} & A_{rr} & A_{r\Om} \\
		  A_{t\Om} & A_{r\Om} & A_{\Om\Om'}
	\end{pmatrix}	\nonumber\\
&=&\sumint\!
	\begin{pmatrix}
		A_{tt} n_L \,&\, A_{tr} n_L \,&\,
			A_{t} \d_{\Om}n_L \!+\! A_{t\Valph} n_{\aleph\Om}^L\\
		\cdots \,&\, A_{rr} n_L \,&\,
			A_{r} \d_{\Om}n_L \!+\! A_{r\Valph} n_{\aleph\Om}^L\\
		\cdots \,&\, \cdots \,&\,
			A_{S} n^L_{\Om\Om'}
				\!+\! \tilde{A}_{S} \tilde{n}^L_{\Om\Om'}
				\!+\! A_\Valph n^{L}_{\AlOm\Om'}
				\!+\! A_{\Talbt} n^{L}_{\albt\Om\Om'}
	\end{pmatrix}
	\! e^{-i\w t},~~~~~
\label{spherical tensor decomposition}
\eea
where we use the scalar multipoles $n^L$, the divergenceless vector multipoles $n^{L}_\AlOm$\footnote
{
The vector multipoles are enumerated by an antisymmetric multi-index $\aleph$ taken from the Hebrew alphabet, representing $D-3$ \emph{spherical} indices:
\bea
n^{L}_\AlOm = \eps^{(\hd+1)}_{\AlOm \, \Om'} \, D^{\Om'} n^L
	= \( \star (\vec{r} \wedge \! \vec\nabla )\)_{\!\AlOm}  \!\!\!\!n^L	~,
\label{x L al Om}
\eea
where $\eps^{(\hd+1)}_{\Om_1\cdots\Om_{\hd+1}}$ is the completely antisymmetric symbol on the $\Omd$-sphere, $\wedge$ is the exterior product and $\star$ is the Hodge duality operator \cite{{{Hestenes1, Hestenes2, Baylis, Doran}}}.
The spatial Levi-Civita tensor will be marked $\eps^{(D)}_{a_1\cdots a_D}$.
} 
and the tensor multipoles $n^{L}_{\aleph\aleph'\Om\Om'}$ (which are symmetric, traceless and divergenceless)\footnote
{
The tensor multipoles (of rank 2, generalizing \cite{Applequist}) are enumerated by 2 antisymmetric multi-indices $\aleph,\aleph'$:
\bea
n^{L}_{\aleph\aleph'\Om\Om'}
	= \eps^{(\hd+1)}_{\aleph \Om \Psi} \eps^{(\hd+1)}_{\aleph' \Om' \Psi'} D^{\Psi} \! D^{\Psi'} \!\!\! n^L
~.
\label{x L al al' Om Om'}
\eea
}. 
These multipoles are all dimensionless, and depend only on the angular coordinates.
They are related to the scalar, vector and tensor spherical harmonics:
\bea
n^L &=& \frac{x^L}{r^\ell} = Y^S ~~,~~
	D_\Om n^L= Y^S_\Om  ~~,~~
	n^L_{\Om\Om'}=g_{\Om\Om'}n^L  ~~,~~
	\tilde{n}^L_{\Om\Om'}=\! \( \! D_\Om D_{\Om'} +\frac{c_s}{\hd\!+\!1} g_{\Om\Om'} \! \) \! n^L ,~~
\nonumber\\
	n^{L}_{\aleph\Om} &=& \frac{x^{L}_{\aleph\Om}}{r^\ell} = Y^\Valph_\Om  ~~,~~
	n^L_{\aleph\Om\Om'}=\frac{1}{2}\( D_\Om n^L_{\aleph\Om'} + D_{\Om'} n^L_{\aleph\Om} \)~,~
	n^{L}_{\albt\Om\Om'} = \frac{x^{L}_{\albt\Om\Om'}}{r^\ell} =  Y^{\Talbt}_{\Om\Om'}.~~~
\label{solid harmonics}
\eea
Using \cite{PaperIII, RubinOrdonez, Higuchi:1986wu}, the numbers $D_\ell(\hd+1,\eps)$ of such independent multipoles of order $\ell$ and sector $\eps$ on the $\Omd$-sphere are
\bea
D_\ell(\hd+1,0) = \frac{(2\ell+\hd)(\ell+\hd-1)!}{\hd! \, \ell!}~~~~,~~~~
D_\ell(\hd+1,1) = \frac{\ell(\ell+\hd)(2\ell+\hd)(\ell+\hd-2)!}{(\hd-1)! \, (\ell+1)!}~,
\nonumber\\
D_\ell(\hd+1,2) = \frac{(\hd+2)(\hd-1)(\ell+\hd+1)(\ell-1)(2\ell+\hd)(\ell+\hd-2)!}{2 \, \hd! \, (\ell+1)!}
~~.~~~~~~~~~~~~
\label{Degeneracies}
\eea
The spherical multipoles satisfy the derivative properties (using appendix \ref{app:spherical harmonics})
\bea
\d_t h_{L\w} &=& -i\w h ~~,~~
	\d_r h_{L\w} = h'_{L\w} ~~~~~,~~
	\d_r n^L_X = 0 ~~~,~~
	\d_r x^L_X = \frac{\ell}{r}x^L_X ~~~,~~
\label{derivatives of h}\\
D_\Om h_{L\w} &=& 0 ~~~~~~~\, ,~~
	D_\Om n^{L}_{\aleph\Om} = 0 ~~~~~~~,~~~
	D_\Om n^{L}_{\albt\Om\Om'} = D_{\Om'} n^{L}_{\albt\Om\Om'} = 0~,
\label{divergenceless}\\
g^{\Om\Om'} n^L_{\aleph\Om\Om'} &=& 0~~~~~~~,~~
	D_\Om \tilde{n}^L_{\Om\Om'} = -\frac{\hd\,\hc_s}{\hd\!+\!1} D_{\Om'}n^L~~,~~
	D_\Om n^L_{\aleph\Om\Om'} = -\frac{\hc_s}{2} n^L_{\aleph\Om'}~,
\label{divergences}
\eea
where a prime denotes r-derivatives ($\,'\!\!:=\d_r$).
They are eigenfunctions of the Laplace-Beltrami operator on the $(\hd+1)$-sphere, $\Lapd=D_\Om D^\Om$, with eigenvalues
\bea
\Lapd n^L &=& -c_s n^L ~,~~
	\Lapd n^L_{\Om\Om'} = -c_s n^L_{\Om\Om'} ~,~~
 	\Lapd D_\Om x^L = -(c_s\!-\!\hd) D_\Om n^L
\label{Laplacians}\\
\Lapd n^{L}_{\aleph\Om} &=& -(c_s\!-\!1) n^{L}_{\aleph\Om} ~~~,~~
	\Lapd n^L_{\aleph\Om\Om'} = -(c_s \!-\! d) n^L_{\aleph\Om\Om'} ~,~
\label{Laplacians2}\\
\Lapd \tilde{n}^L_{\Om\Om'}&=& -(c_s-2(\hd\!+\!1)) \tilde{n}^L_{\Om\Om'} ~,~
	\Lapd n^{L}_{\albt\Om\Om'} = -(c_s\!-\!2) n^{L}_{\albt\Om\Om'}
\label{Laplacians3}.
\eea
The scalar, vector and tensor basis elements are orthogonal to each other, and we use the following normalization conditions in d dimensions (\cite{RubinOrdonez,Higuchi:1986wu}),
\bea
\int \!\!\dOmd n_{L_\ell}(\Omd) n^{L'_{\ell '}}(\Omd)
&=&
\Nld \Omd \delta_{\ell \ell'} \delta_{L_\ell L'_{\ell'}} \, , \nonumber \\
\int \!\!\dOmd g^{\Om \Om'} D_{\Om} n_{L_\ell} D_{\Om'} n^{L'_{\ell '}}
&=&
c_s \cdot \Nld \Omd \delta_{\ell \ell'} \delta_{L_\ell L'_{\ell'}} \, , \nonumber \\
\int \!\!\dOmd g^{\Om \Om'} n^{L_\ell}_{\aleph \Om} n^{L'_{\ell'}}_{\aleph' \Om'}
&=&
c_s \cdot \Nld \Omd \delta_{\ell \ell'} \delta_{L_\ell L'_{\ell'}} \delta_{\aleph\aleph'}\, ,\nonumber \\
\int \!\!\dOmd g^{\Om \Om'} \! g^{\Psi \Psi'} \!\! n^{L_\ell}_{\Om\Psi} n^{L'_{\ell'}}_{\Om'\Psi'}
&=&
\VERBOSE{	
\!\! \int \!\!\dOmd n^{L'_{\ell'}} n^{L_\ell}
	g^{\Om \Om'} \! g^{\Psi \Psi'} \!\! g_{\Om\Psi} g_{\Om'\Psi'}
	\!=\!
}	
(\hd\!+\!1) \Nld \Omd \delta_{\ell \ell'} \delta_{L_\ell L'_{\ell'}} , \nonumber\\
\int \!\!\dOmd g^{\Om \Om'} \! g^{\Psi \Psi'} \! \tilde{n}^{L_\ell}_{\Om\Psi} \tilde{n}^{L'_{\ell'}}_{\Om'\Psi'}
&=&
\VERBOSE{	
\!\! \int \!\!\dOmd g^{\Om \Om'} \! g^{\Psi \Psi'} \!\!
	\( \!\! D_{\Om}D_{\Psi} \!+\! \frac{c_s}{\hd\!+\!1}g_{\Om\Psi} \!\!\) \! n^{L_\ell} \!\!
	\(\!\!D_{\Om'}D_{\Psi'}\!+\!\frac{c'_s}{\hd\!+\!1}g_{\Om'\Psi'}\!\!\) \! n^{L'_{\ell'}}
\nonumber\\
&=& \!\! \int \!\!\dOmd g^{\Om \Om'} \! g^{\Psi \Psi'} \!\!
\[	D_{\Om}D_{\Psi}n^{L_\ell} D_{\Om'}D_{\Psi'}n^{L'_{\ell'}}
	+ \frac{c_s c'_s}{(\hd\!+\!1)^2}g_{\Om\Psi}n^{L_\ell}g_{\Om'\Psi'}n^{L'_{\ell'}}
\right.\nonumber\\
&&~~~~~~~~~~~~~~
\left.
	+ \frac{c_s}{\hd\!+\!1}g_{\Om\Psi}n^{L_\ell} D_{\Om'}D_{\Psi'}n^{L'_{\ell'}}
	+ D_{\Om}D_{\Psi}n^{L_\ell} \frac{c'_s}{\hd\!+\!1}g_{\Om'\Psi'}n^{L'_{\ell'}}
\]
\nonumber\\
&=& \!\! \int \!\!\dOmd \!\!
\(	- g^{\Psi \Psi'} \Lapd D_{\Psi}n^{L_\ell} D_{\Psi'}n^{L'_{\ell'}}
	+ \frac{c_s c'_s}{\hd\!+\!1}n^{L_\ell}n^{L'_{\ell'}}
\right.\nonumber\\
&&~~~~~~~~~
\left.
	+ \frac{c_s}{\hd\!+\!1} n^{L_\ell} \Lapd n^{L'_{\ell'}}
	+ \frac{c'_s}{\hd\!+\!1} n^{L'_{\ell'}} \Lapd n^{L_\ell}
\)
\nonumber\\
\VERBOSEE{	
&=& \!\! \int \!\!\dOmd \!\!
\(	- (c_s-\hd) n^{L'_{\ell'}} g^{\Psi \Psi'} D_{\Psi'} D_{\Psi}n^{L_\ell}
	- \frac{c_s c'_s}{\hd\!+\!1}n^{L_\ell}n^{L'_{\ell'}}
\)
\nonumber\\
&=& \!\! \int \!\!\dOmd n^{L_\ell}n^{L'_{\ell'}} \( c_s (c_s-\hd) - \frac{c_s c'_s}{\hd\!+\!1} \)
\nonumber\\
}	
&=&
}	
\frac{\hd \, c_s \, \hc_s}{\hd\!+\!1}
	\cdot \Nld \Omd \delta_{\ell \ell'} \delta_{L_\ell L'_{\ell'}} \, , \nonumber\\
\int \!\!\dOmd  g^{\Om \Om'} \! g^{\Psi \Psi'} \!\! n^{L_\ell}_{\AlOm\Psi} n^{L'_{\ell'}}_{\aleph'\Om'\Psi'}
&=&
\VERBOSE{	
\frac{1}{4}\int \!\!\dOmd g^{\Om \Om'} \! g^{\Psi \Psi'} \!\!
	\(D_{\Om}n^{L_\ell}_{\aleph\Psi}\!+\!D_{\Psi}n^{L_\ell}_{\aleph\Om}\)
	\(D_{\Om'}n^{L'_{\ell'}}_{\aleph'\Psi'}\!+\!D_{\Psi'}n^{L'_{\ell'}}_{\aleph'\Om'}\)
\nonumber\\
\VERBOSEE{	
&=& \frac{1}{4}\int \!\!\dOmd \!\!
\[	g^{\Om \Om'} \! g^{\Psi \Psi'} D_{\Om}n^{L_\ell}_{\aleph\Psi}D_{\Om'}n^{L'_{\ell'}}_{\aleph'\Psi'}
	+g^{\Om \Om'} \! g^{\Psi \Psi'} D_{\Psi}n^{L_\ell}_{\aleph\Om}D_{\Om'}n^{L'_{\ell'}}_{\aleph'\Psi'}
\right.\nonumber\\
&&~~~~~~~~~~~~
\left.
	+g^{\Om \Om'} \! g^{\Psi \Psi'}D_{\Om}n^{L_\ell}_{\aleph\Psi}D_{\Psi'}n^{L'_{\ell'}}_{\aleph'\Om'}
	+g^{\Om \Om'} \! g^{\Psi \Psi'}D_{\Psi}n^{L_\ell}_{\aleph\Om}D_{\Psi'}n^{L'_{\ell'}}_{\aleph'\Om'}
\]
\nonumber\\
}	
&=& -\frac{1}{2}\int \!\!\dOmd g^{\Psi \Psi'}
\[	n^{L_\ell}_{\aleph\Psi} \Lapd n^{L'_{\ell'}}_{\aleph'\Psi'}
	+n^{L'_{\ell'}}_{\aleph'\Psi'}g^{\Om \Om'} D_{\Om'} D_{\Psi}n^{L_\ell}_{\aleph\Om}
\]
\nonumber\\
&=& -\frac{1}{2}\int \!\!\dOmd g^{\Psi \Psi'}
\[	-(c_s\!-\!1) n^{L_\ell}_{\aleph\Psi} n^{L'_{\ell'}}_{\aleph'\Psi'}
	+n^{L'_{\ell'}}_{\aleph'\Psi'}g^{\Om \Om'}
		[D_{\Om'},D_{\Psi}]n^{L_\ell}_{\aleph\Om}
\]
\nonumber\\
&=& -\frac{1}{2}\int \!\!\dOmd g^{\Psi \Psi'}
\[	-(c_s\!-\!1) n^{L_\ell}_{\aleph\Psi} n^{L'_{\ell'}}_{\aleph'\Psi'}
	+\hd n^{L'_{\ell'}}_{\aleph'\Psi'} n^{L_\ell}_{\aleph\Psi}
\]
\nonumber\\
&=&
}	
\frac{c_s \, \hc_s}{2} \cdot \Nld \Omd \delta_{\ell \ell'} \delta_{L_\ell L'_{\ell'}} \, ,
\nonumber\\
\int\!\!\dOmd g^{\Om \Om'}g^{\Psi \Psi'}n^{L_\ell}_{\albt\Om\Psi}
	n^{L'_{\ell'}}_{\aleph'\beth'\Om'\Psi'}
&=&
\TensorNorm\!\cdot \Nld \Omd \delta_{\ell \ell'} \delta_{L_\ell L'_{\ell'}} \delta_{\aleph\aleph'}
	\delta_{\beth\beth'} \, .
\label{sphercal harmonics normalization}
\eea

\subsection{More spherical expressions}
The $d$-dimensional trace of $A_{\mu\nu}$ is given by
\bea
A\!\!&:=&\!\!g^{\al\altld} A_{\al\altld} \!=\!
\VERBOSE{	
\!\sumint\!\[\! -A_{tt}n^L \!\!+\!\! A_{rr}n^L
	\!\!+\!g_d^{\Om\Om'}\!\!
		\( \! A_{S} n^L_{\Om\Om'}
			\!\!+\!\! \tilde{A}_{S} \tilde{n}^L_{\Om\Om'}
			\!\!+\!\! A_\Valph n^{L}_{\AlOm\Om'}
			\!\!+\!\! A_{\Talbt} n^{L}_{\albt\Om\Om'}
		\!\)
	\!\] \!\! e^{-i\w t} \nonumber\\
&=&
}	
\!\sumint\( -A_{tt} + A_{rr} + \frac{\hd+1}{r^2} A_{S} \) n^L e^{-i\w t}
\triangleq \!\sumint A \, n^L \, e^{-i\w t},
\label{trace A}
\eea
and its divergence is given by
\bea
(\text{div} A)_\al &=& A_{\al\bt;}\!\,^{\bt}=
\VERBOSE{	
	-D_t A_{t\al} +D_r A_{r\al} +g_d^{\Om\Om'} D^d_{\Om'}A_{\al\Om}
	\nonumber\\
&=&
}	
\!-\d_t A_{t\al} \!+\! \( A_{r\al,r}\!-\!\Gm^{\mu}_{r\al}A_{\mu r} \)
\!+\!\frac{g^{\Om\Om'}}{r^2}\!
	\(A_{\al\Om,\Om'}\!-\!\Gm^{\mu}_{\Om'\al}A_{\mu\Om}\!-\!\Gm^{\mu}_{\Om\Om'}A_{\mu\al}\)
\!.~~~~~~~~
\label{spherical divergence}
\eea
Explicitly, its components are given by
\bea
(\text{div} A)_t &=&
\VERBOSE{	
	-\d_t A_{tt} +A_{tr,r}
	+\frac{g^{\Om\Om'}}{r^2} \(A_{t\Om,\Om'} -\Gm^{r}_{\Om\Om'}A_{tr}
						-\Gm^{\Om''}_{\Om\Om'}A_{t\Om''}\)
	\nonumber\\
&=& -\d_t A_{tt} +A_{tr,r} + \frac{g^{\Om\Om'}}{r} g_{\Om\Om'}A_{tr}
	+\frac{g^{\Om\Om'}}{r^2} D_{\Om'}A_{t\Om}
	\nonumber\\
&=& \sumint \! e^{-i\w t} \! \[ i\w A^{L\w}_{tt}n_L
					\!+\!\( \d_r \!+\! \frac{\hd+1}{r} \) \!\! \( A^{L\w}_{tr} n_L \)
					\!+\!\frac{g^{\Om\Om'}}{r^2} D_{\Om'}
						\( A^{L\w}_{tS} D_\Om n_L \!
						+\! A^{L\w}_{t\Valph}  n^L_{\aleph\Om} \)
				\]	\nonumber\\
&=& \sumint \! e^{-i\w t} \! \[ i\w A^{L\w}_{tt}n_L
					+n_L \( \d_r + \frac{\hd+1}{r} \) \!\! A^{L\w}_{tr}
					+\frac{1}{r^2}A^{L\w}_{tS}\Delta_{\Omd} n_L
				\]	\nonumber\\
&=&
}	
\sumint \! e^{-i\w t} \! \[ i\w A_{tt}
					+\( \d_r + \frac{\hd+1}{r} \) \!\! A_{tr}
					-\frac{c_s}{r^2}A_{t}
				\] n_L,
\label{spherical divergence t}
\eea
\bea
(\text{div} A)_r &=&
\VERBOSE{	
	-\d_t A_{tr} +A_{rr,r}
	+\frac{g^{\Om\Om'}}{r^2}
		\(A_{r\Om,\Om'}
			-\Gm^{\Om''}_{r\Om'}A_{\Om\Om''}
			-\Gm^{r}_{\Om\Om'}A_{rr}
			-\Gm^{\Om''}_{\Om\Om'}A_{r\Om''}
		\)	\nonumber\\
&=& -\d_t A_{tr} +A_{rr,r}
	+\frac{g^{\Om\Om'}}{r^2}
		\(D_ {\Om'} A_{r\Om}
			-\frac{1}{r}g^{\Om''}_{\Om'}A_{\Om\Om''}
			+r g_{\Om\Om'}A_{rr}
		\)	\nonumber\\
&=& -\d_t A_{tr} +A_{rr,r}
	+\frac{1}{r^2}
		\(g^{\Om\Om'} D_ {\Om'} A_{r\Om}
			-\frac{\hd+1}{r}A_S
			+(\hd+1)r A_{rr}
		\)	\nonumber\\
&=&
}	
\sumint \! e^{-i\w t} \! \[ i\w A_{tr} + \(\d_r + \frac{\hd+1}{r} \) A_{rr}
					-\frac{c_s}{r^2}A_{r} -\frac{\hd+1}{r^3}A_{S}
				\] n_L,~~
\label{spherical divergence r}
\eea
\bea
(\text{div} A)_{\Om} &=&
\VERBOSE{	
-\d_t A_{t\Om} +A_{r\Om,r} -\Gm^{\Om'}_{r\Om}A_{r\Om'}	\nonumber\\
&&+\frac{g^{\Om''\Om'}}{r^2}
	\(A_{\Om\Om'',\Om'}
		-\Gm^{r}_{\Om'\Om}A_{r\Om''} -\Gm^{\Om'''}_{\Om'\Om}A_{\Om'''\Om''}
		-\Gm^{r}_{\Om''\Om'}A_{r\Om} -\Gm^{\Om'''}_{\Om''\Om'}A_{\Om'''\Om}
	\)	\nonumber\\
&=& -\d_t A_{t\Om} +A_{r\Om,r} -\frac{1}{r}g^{\Om'}_{\Om}A_{r\Om'}
+\frac{g^{\Om''\Om'}}{r^2}
	\( D_{\Om'} A_{\Om\Om''} +r g_{\Om'\Om}A_{r\Om''} +r g_{\Om''\Om'}A_{r\Om}
	\)	\nonumber\\
&=& -\d_t A_{t\Om} +A_{r\Om,r} -\frac{1}{r}A_{r\Om}
+\frac{1}{r^2}
	\( D^{\Om''} A_{\Om\Om''} +r A_{r\Om} +(\hd+1)rA_{r\Om}
	\)	\nonumber\\
&=& -\d_t A_{t\Om} +\(\d_r+\frac{\hd+1}{r}\) \! A_{r\Om}
	+\frac{1}{r^2} \( D^{\Om''} A_{\Om\Om''} \)	\nonumber\\
&=&
}	
\sumint \! e^{-i\w t} \! \left\{ \[
			i\w A_{t}
			\!+\!\(\d_r\!+\!\frac{\hd\!+\!1}{r}\) \! A_{r}
			\!+\! \frac{A_{S}}{r^2}
			-\!\frac{\hd \, \hc_s}{(\hd\!+\!1) \, r^2} \tilde{A}_{S}
		\] \right. \!\! D_\Om n_L \nonumber\\
&&~~~~~~~~~~+\!\!\left. \[
			i\w A_{t\Valph}
			+\(\d_r\!+\!\frac{\hd\!+\!1}{r}\) \! A_{r\Valph}
			- \frac{\hc_s}{2 r^2} A_\Valph
		\] \! n^L_{\aleph\Om} \right\}
\label{spherical divergence Om},
\eea
where we have used
\bea
D^{\Om'} A_{\Om\Om'}&=&
\VERBOSE{	
\sumint \! e^{-i\w t} \! \[
				A_{S} D^{\Om'} n^L_{\Om\Om'}
				+ \tilde{A}_{S} D^{\Om'}\tilde{n}^L_{\Om\Om'}
				+ A_\Valph D^{\Om'} n^{L}_{\AlOm\Om'}
				+ A_{\Talbt} D^{\Om'} n^{L}_{\albt\Om\Om'}
			\] ~~~~\nonumber\\
&=&
}	
\sumint \! e^{-i\w t} \! \[
	\( A_{S} -\frac{\hd \, \hc_s}{\hd\!+\!1} \tilde{A}_{S} \) \! D_{\Om}n^L
	-\frac{\hc_s}{2} n^L_{\aleph\Om} A_\Valph
	\]\!.~~~~~~~
\eea

\section{Action in spherical fields: Kinetic terms}
\label{app:Homogenous action in spherical fields}
In this appendix we express the linearized GR kinetic term with gauge-invariant spherical fields.
We start with the action (\ref{S_hom}).
\VERBOSE{	
\bea
\Sh
&=& \ShPreFhalf \int \!\! d^d x \sqrt{-g} \, g^{\al\altld} g^{\bt\bttld} g^{\gm\gmtld}
	\[\frac{1}{2}(D_\gm h_{\al[\altld}) (h_{\bt]\bttld;\gmtld})
	  +(D_{\bttld} h_{\gm[\bt}) (h_{\gmtld]\al;\altld})
	 \]	\nonumber\\
&=& S_1-S_2+S_3-S_4,	\nonumber\\
S_1&=& \ShPreFquarter \int \!\! d^d x \sqrt{-g} \, g^{\al\altld} g^{\bt\bttld} g^{\gm\gmtld}
	\[(D_\gm h_{\al\altld}) (h_{\bt\bttld;\gmtld})
	 \],		\nonumber\\
S_2&=& \ShPreFquarter \int \!\! d^d x \sqrt{-g} \, g^{\al\altld} g^{\bt\bttld} g^{\gm\gmtld}
	\[(D_\gm h_{\al\bt}) (h_{\altld\bttld;\gmtld})
	 \],		\nonumber\\
S_3&=& \ShPreFhalf \int \!\! d^d x \sqrt{-g} \, g^{\al\altld} g^{\bt\bttld} g^{\gm\gmtld}
	\[ (D_{\bttld} h_{\gm\bt}) (h_{\gmtld\al;\altld})
	 \],		\nonumber\\
S_4&=& \ShPreFhalf \int \!\! d^d x \sqrt{-g} \, g^{\al\altld} g^{\bt\bttld} g^{\gm\gmtld}
	\[(D_{\bttld} h_{\gm\gmtld}) (h_{\bt\al;\altld})
	 \].
\eea
}	
Note that the terms containing traces of the perturbation $\sim D h^{\alpha}_\alpha$ can be expressed only with $d$-scalars, and thus do not contribute to any vectorial or tensorial terms; and that in fact tensorial terms arise only from the term $\sim D^{\gamma} h^{\alpha \beta} \, D_{\gamma} h_{\alpha \beta}$.
After expansions, partial integrations and use of the normalization conditions, we find the homogenous part of the action ($L,\w$ indices will be suppressed throughout much of this appendix),
\be
\Sh
\VERBOSE{	
= \sumint S^{L\w}_{L\w}
= \sumintexplicit \int \!\! dr L^{L\w}_{L\w}(r)
}	
= \sumintexplicit \int \!\! dr L(r)
= \sumintexplicit \ShPreFCanon \int \!\! r^{\hd+1}dr \tilde{L}(r),
\label{homogenous action}
\ee
where the action decomposes into scalar, vector and tensor sectors
\be
\tilde{L}(r)=\tilde{L}_{S}(r)+\tilde{L}_{V}(r)+\tilde{L}_{T}(r).
\label{Action decomposition}
\ee
The tensor sector includes a single contribution,
\bea
\tilde{L}_{T}&=& h^{*}_{\Talbt} E_{\Talbt},~~~~~~\nonumber\\
E_{\Talbt}&=&
	\frac{\TensorNorm}{ 2 \, r^4}\( \w^2
					\!+\! \d_r^2
					\!+\!\frac{\hd\!-\!3}{r}\d_r
					\!-\!\frac{c_s\!+\!2(\hd\!-\!2)}{r^2}
				\) \! h_{\Talbt}.~~~~~~~~
\label{Tensor Action original variables}
\eea
This matches (5.6) of \cite{AsninKol}, if we set $\Phi_{\Talbt} \sim r^{-2}h_{\Talbt}$. Defining
\be
\fieldh_{\Talbt}=r^{-(\ell+2)}h_{\Talbt},
\label{field hT homogenous}
\ee
it can also be presented in the master-form
\bea
L_{T} = \ShPreFCanon \frac{\TensorNorm}{2} \, r^{2\ell+\hd+1} \,
	\fieldh^{*}_{\Talbt} \, \IL \, \fieldh_{\Talbt},
\label{Tensor Action Master Homogenous}
\eea
where $\IL$ is the master operator defined in (\ref{Master wave operator}).

The vector sector is comprised of three terms,
\bea
\tilde{L}_{V}&=& h^{*}_{t\Valph} E_{t\Valph}+h^{*}_{r\Valph} E_{r\Valph}
	+h^{*}_{\Valph} E_{\Valph},	\\
E_{t\Valph}
\VERBOSE{	
&=&-\frac{c_s}{r^2}
	\[ \blue{ \( \Box h_{t\Valph} \) }
	\purple{ +i\w \!\(\!\!
			i\w h_{t\Valph}
			\!+\!\(\d_r\!+\!\frac{\hd\!\!+\!\!1}{r}\) \!\! h_{r\Valph}
			\!-\! \frac{\hc_s}{2 r^2} h_\Valph
		\! \)\!\!	}
	\]\nonumber\\
\VERBOSEE{	
&=&-\frac{c_s}{r^2}
	\[ \blue{ \( \w^2 +\d_r^2+\frac{\hd\!-\!1}{r}\d_r-\frac{c_s\!+\!\hd\!-\!1}{r^2} \)
		h_{t\Valph} }
	\right.\nonumber\\
&&~~~~~~\left.
	\purple{ +i\w \!\(\!\!
			i\w h_{t\Valph}
			\!+\!\(\d_r\!+\!\frac{\hd\!\!+\!\!1}{r}\) \!\! h_{r\Valph}
			\!-\! \frac{\hc_s}{2 r^2} h_\Valph
		\! \)\!\!	}
	\]\nonumber\\
}	
}	
&=&-\frac{c_s}{r^2}
	\[ \! \blue{ \(\!\! \d_r^2\!+\!\frac{\hd\!\!-\!\!1}{r}\d_r\!-\!\frac{c_s\!+\!\hd\!-\!1}{r^2}\! \) \!\!
			h_{t\Valph} }
	\purple{ +i\w \!\(\!\!
			\(\! \d_r\!+\!\frac{\hd\!\!+\!\!1}{r} \! \) \!\! h_{r\Valph}
			\!-\! \frac{\hc_s}{2 r^2} h_\Valph
		\! \)\!	}
	\]\!\!,~~~~~~~~
\label{EtV}\\
E_{r\Valph}
\VERBOSE{	
&=& \frac{c_s}{r^2}
	\[ \blue{\( \Box h_{r\Valph}^{L\w} \) }
	    \purple{ -\(\d_r\!-\!\frac{2}{r}\) \!
		\! \(\!
			i\w h^{L\w}_{t\Valph}
			\!+\!\(\d_r\!+\!\frac{\hd\!\!+\!\!1}{r}\) \!\! h^{L\w}_{r\Valph}
			\!-\! \frac{\hc_s}{2 r^2} h^{L\w}_\Valph
		\! \) }
	\] \nonumber\\
\VERBOSEE{	
&=& \frac{c_s}{r^2}
	\[ \blue{ \( \w^2 \!\!+\!\d_r^2 \!+\! \frac{\hd\!-\!1}{r}\d_r
				\!-\!\frac{c_s\!+\!2(\hd\!+\!1)}{r^2}
			\) \!\! h^{L\w}_{r\Valph}
		 +\frac{\hc_s}{r^3} h^{L\w}_\Valph }
	\right.\nonumber\\
&&~~~~~~\left.
	    \purple{ -\(\d_r\!-\!\frac{2}{r}\) \!
		\! \(\!
			i\w h^{L\w}_{t\Valph}
			\!+\!\(\d_r\!+\!\frac{\hd\!\!+\!\!1}{r}\) \!\! h^{L\w}_{r\Valph}
			\!-\! \frac{\hc_s}{2 r^2} h^{L\w}_\Valph
		\! \) }
	\]\nonumber\\
}	
}	
&=& \frac{c_s}{r^2}\!
	\[ \! \( \! \w^2 \!\!-\!\frac{\hc_s}{r^2} \! \) \! h_{r\Valph}
		\purple{ -i\w \! \(\!\! \d_r\!-\!\frac{2}{r}\!\) \! h_{t\Valph}}
		\!+\!\frac{\hc_s}{2 r^2}\!\!\(\!\! \d_r\!-\!\frac{2}{r}\!\) \! h_\Valph
	\]\!\!,~~~~~~
\label{ErV}\\
E_\Valph
\VERBOSE{	
&=&\frac{\hc_s}{4\,r^4}
	\[	\blue{ c_s\(\Box h^{L\w}_\Valph\)		}
		\purple{
			-2i\w c_s h^{L\w}_{t\Valph}
			\!-\!2c_s\!\(\!\d_r\!+\!\frac{\hd\!\!+\!\!1}{r}\!\) \!\! h^{L\w}_{r\Valph}
			\!+\!\frac{c_s\hc_s}{r^2} h^{L\w}_\Valph }
	\]\nonumber\\
}	
&=&\frac{c_s\hc_s}{4\,r^4}\!
	\[
		\[ 	\blue{ \frac{4}{r} }
			\purple{-2\!\(\!\d_r\!+\!\frac{\hd\!\!+\!\!1}{r}\!\)} \!\!
		\] \!\! h_{r\Valph}
		\purple{-2i\w h_{t\Valph}	}
		\!+\!\!\[ \blue{ \w^2 \!+\! \d_r^2 \!+\!\frac{\hd\!\!-\!\!3}{r}\d_r }
			\!+\!2\frac{2\!\!-\!\!\hd}{r^2}
		\] \!\! h_{\Valph}
	\]\!\!,~~~~~~~~
\label{EV}
\eea
The field $h_{r\Valph}$ appears in the equation $E_{r\Valph}=0$ (\ref{ErV}) without derivatives.
Thus we can solve for it algebraically,
\be
h_{r\Valph} =	\frac{1}{\Mv(r)}
	\[ i\w r^2 \(\!\! \d_r\!-\!\frac{2}{r}\!\) \! h_{t\Valph}
	-\frac{\hc_s}{2}\!\(\!\! \d_r\!-\!\frac{2}{r}\!\) \! h_\Valph
	\],
\label{homogenous hrV}
\ee
where $\Mv(r)=\w^2r^2 - \hc_s$ (compare (5.9, 5.10) of \cite{AsninKol}).
We thus see $h_{r\Valph}$ is an auxiliary field.
We plug this expression in the equations for $E_{t\Valph}$, $E_{\Valph}$, $L_{V}$,
\VERBOSEE{	
marking $A=i\w r^2 \(\d_r\!-\!\frac{2}{r}\)$, $B=-\frac{\hc_s}{2}\(\d_r\!-\!\frac{2}{r}\)$,
}	
\VERBOSE{	
and calculate (with $C=\d_r\!+\!\frac{\hd+1}{r}$)
\bea
C h^{L\w}_{r\Valph}
&=&
\Mv^{-2} \!
\[ 	\( \w^2r^2 \!-\! \hc_s \) \! i\w r^2 \d_r^2 h^{L\w}_{t\Valph}
	+\( (\hd\!\!-\!\!1) \w^2 r^2 \!-\! \hc_s(\hd\!\!+\!\!1) \) \!  i\w r \d_r h^{L\w}_{t\Valph}
	\right. \nonumber\\
&&~~~~~
	+\(\! \hc_s(\hd\!\!+\!\!2) \!-\! \hd\w^2r^2 \) \!   2i\w h^{L\w}_{t\Valph}
	+\( \hc_s \!-\! \w^2r^2 \) \! \frac{\hc_s}{2} \d_r^2 h^{L\w}_\Valph
	\nonumber\\
&&~~~~~\left.
	+\( \! (\hd\!\!-\!\!1)\hc_s \!-\! (\hd\!\!-\!\!3)\w^2r^2 \) \!\! \frac{\hc_s}{2r} \d_r  h^{L\w}_\Valph
	+\( \! (\hd\!\!-\!\!2) \w^2r^2 \!-\! \hd \hc_s \) \!\! \frac{\hc_s}{r^2} h^{L\w}_\Valph
\]\!,~~~~~~~~
\eea
}	

\bea
\tilde{L}_{V}&=&\blue{h^{*}_{t\Valph} E_{t\Valph}}
	\purple{+h^{*}_{\Valph} E_{\Valph}}	\nonumber\\
\VERBOSE{	
&=&
\blue{-2\frac{c_s}{r^2} h^{L\w*}_{t\Valph} \!
	\[ \! \(\!\! \d_r^2\!+\!\frac{\hd\!\!-\!\!1}{r}\d_r\!-\!\frac{c_s\!+\!\hd\!-\!1}{r^2}\! \) \!\!
			h_{t\Valph}^{L\w}
		- \frac{i\w \hc_s}{2 r^2} h^{L\w}_\Valph
		+ i\w C h^{L\w}_{r\Valph}
	\]	}
	\nonumber\\
&&
\purple{+\frac{c_s \hc_s}{2\,r^4} h^{L\w*}_{\Valph}  \!
	\[
		\[ \w^2 \!+\! \d_r^2 \!+\!\frac{\hd\!-\!3}{r}\d_r \!+\!2\frac{2\!-\!\hd\!}{r^2}
			\] \!\! h^{L\w}_{\Valph}
		-2i\w h^{L\w}_{t\Valph}
		+ \[ \frac{4}{r} -2 C \] \! h_{r\Valph}^{L \w}
	\]	}	\nonumber\\
&=&
\blue{h^{*}_{t\Valph} \frac{2 c_s \hc_s}{r^2 \, \Mv}
\[
	\d_r^2
	+ \(  \hd \!-\! 3 -\frac{2\hc_s}{\Mv} \) \! \frac{1}{r}\d_r
	+ \(2 - \hd + \frac{2\hc_s}{\Mv} +\frac{\Mv}{2}	\) \!\frac{2}{r^2}
\] h_{t\Valph} }
	\nonumber\\
&&
\blue{+h^{*}_{t\Valph} \, \frac{i\w c_s \hc_s}{r^2\,\Mv}
	\[	\d_r^2
		+\(\hd\!-\!3 \!-\! \frac{2\hc_s}{\Mv} \) \! \frac{1}{r} \d_r
		+\!\( 2-\hd \!+\! \frac{2\hc_s}{\Mv} \!+\!\frac{\Mv}{2} \)\!\frac{2}{r^2}
	\]  h_\Valph }		\nonumber\\
&&
\purple{-h^{*}_{\Valph} \, \frac{i\w c_s \hc_s}{r^2\,\Mv}
	\[	\d_r^2
		+\(\hd\!-\!3 \!-\! \frac{2\hc_s}{\Mv} \) \! \frac{1}{r} \d_r
		+\!\( 2-\hd \!+\! \frac{2\hc_s}{\Mv} \!+\!\frac{\Mv}{2} \)\!\frac{2}{r^2}
	\] h_{t\Valph} }		\nonumber\\
&&
\purple{+h^{*}_{\Valph} \, \frac{c_s \hc_s \,\w^2}{2\,r^2\,\Mv} \!
	\[
		\d_r^2
		+\(\hd\!-\!3\!-\!\frac{2\hc_s}{\Mv}\! \) \!
			\frac{1}{r}\d_r
		\!+\! \( \!  2-\hd \!+\! \frac{2\hc_s}{\Mv}\!+\! \frac{\Mv}{2}\!\)\!\!
		\frac{2}{r^2}
	\] h_\Valph }	\nonumber\\
}	
&=&
\frac{c_s \hc_s}{4 \, r^2 \, \Mv}
	\( r^2 \Phi_{\Valph} \)^{\!*} \! \IL_V \( r^2 \Phi_{\Valph} \)
=
\frac{c_s \hc_s}{4} \Phi^{*}_{\Valph} \!
	\[ \Phi_{\Valph}
		+ \frac{1}{r^{\hd+1}}\d_r \! \( \! \frac{r^{\hd+3}}{\Mv}	\d_r \Phi_{\Valph} \)
	\!\] \!,~~~~~~~~
\label{Vector Action}
\eea
where $\IL_V=	\d_r^2
			+\! \(\!\hd\!-\!3\!-\!\frac{2\hc_s}{\Mv}\! \) \! \frac{1}{r}\d_r
			\!+\! \( \!  2\!-\!\hd \!+\! \frac{2\hc_s}{\Mv}\!+\! \frac{\Mv}{2}\!\)\!\!
			\frac{2}{r^2}$,
and the vector part of the action is seen to depend only on the \emph{single} gauge-invariant field (compare (5.12, 5.13) of \cite{AsninKol}, and \cite{KodamaIshibashi})
\be
\Phi_{\Valph} = r^{-2} \(2 h_{t\Valph}\!+\! i\w h_{\Valph} \).
\label{Vector Field}
\ee
Since $\tilde{L}_{V}$ only appears in the integral (\ref{homogenous action}), we can integrate by parts to obtain
\be
\tilde{L}_{V}=
\frac{c_s \hc_s}{4} \!
	\[ \(\Phi_{\Valph}\)^2 - \frac{r^2}{\Mv(r)} \( \d_r \Phi_{\Valph} \)^2 \].
\label{Homogenous Vector Action by parts}
\ee
Defining the field $\fieldh_{\Valph}$ using the canonical transformations
\bea
\( 2 (\ell\!-\!1) (\ell\!+\!\hd) \, r^{\ell+\hd+1} \fieldh_{\Valph} \) \!
	:=\frac{\partial L}{\partial\!\(\Phi^{*}_{\Valph}\)'}
	&=&-\frac{c_s \hc_s}{2\Mv} r^{\hd+3} \! \(\Phi_{\Valph}\)'	\nonumber\\
&\ergo&
	\(\Phi_{\Valph}\)' =
		-\frac{4 \, r^{\ell}}{\ell (\ell\!+\!\hd\!+\!1)} \! \( \! \w^2 \! - \! \frac{\hc_s}{r^2} \! \) \! \fieldh_{\Valph}\,,~~~~~~~~~~
\label{field hV homogenous 1}
\\
\( 2 (\ell\!-\!1) (\ell\!+\!\hd) \, r^{\ell+\hd+1} \fieldh_{\Valph} \) ' \!\!
	:=\frac{\partial L}{\partial\!\(\Phi^{*}_{\Valph}\)}
	&=& \half c_s \hc_s r^{\hd+1} \Phi_{\Valph}		\nonumber\\
&\ergo&
	\Phi_{\Valph} = \frac{4}{\ell (\ell\!+\!\hd\!+\!1) \, r^{\hd+1}} \( r^{\ell+\hd+1} \fieldh_{\Valph} \) ' \!,~~~~~~~~~~
\label{field hV homogenous 2}
\eea
and again using integration by parts, we can present the vector contribution as well in the master-form (compare with (\ref{Tensor Action Master Homogenous}), differing only in the prefactor constants)
\bea
L_{V} = \ShPreFCanon \frac{8 (\ell\!-\!1) (\ell\!+\!\hd)}{(\ell\!+\!\hd\!+\!1) \ell} r^{2\ell+\hd+1} \,
	\fieldh^{*}_{\Valph} \, \IL \, \fieldh_{\Valph}.~~~~~~~~
\label{Vector Action Master Homogenous}
\eea

In the remaining scalar sector, we also expect a reduction to a single gauge-invariant field.
We first notice that 3 of the fields ($h_{tr}$, $h_{rr}$, $h_{r}$) appear without $r$-derivatives in the action; thus variations with respect to their complex conjugates ($h^{*}_{tr}$, $h^{*}_{rr}$, $h^{*}_{r}$) give us algebraic equations\footnote{
The complete derivation of the scalar sector of the action, along with the collection of terms to the gauge invariant combination, can be found in Mathematica code in our additional notes online \cite{Code}.
}.
Hence these fields can be found by solving the system given by the (hermitian) matrix $A$\footnote{
We mark that this matrix is identical to (4.35) of \cite{AsninKol}, adapted to a flat background - up to ordering of the fields, scalings in the field definitions, and sign mismatches stemming from the choice of Lorentzian over Euclidean signature.
Using our definitions, the matrices $A$, $A^{-1}$, $\tilde{A}^{-1}$ are dimensionless.
},
\bea
A
\(\begin{matrix}
	h_{tr}\\
	h_{rr}\\
	\frac{c_s}{r}h_{r}
 \end{matrix}\)
=B;~~
A :=
\(\begin{matrix}
  c_s 			&	 -(\hd+1) i\w r 			&~	 i\w r						\\
  (\hd+1) i\w r 	&	 \frac{\hd(\hd+1)}{2} 	&~	-\hd	 						\\
  -i\w r			&	 -\hd		 			&~	 \frac{\w^2r^2 + 2\hd}{c_s}	\,
\end{matrix}\), ~~
B=
	\(\begin{matrix}
		B_1\\ B_2 \\ B_3,
	 \end{matrix}\),~~~
\label{Matrix A}
\eea
with 
\bea
B_1 &:=&
	c_s \d_r h_{t}
	- (\hd\!\!+\!\!1) i\w \! \( \! \d_r \!-\!\frac{1}{r} \! \) \!\! h_{S}
	\nonumber\\
B_2 &:=&
	\frac{c_s \!-\! (\hd\!+\!1) r\d_r}{2} h_{tt}
	\!+\! i\w c_s h_{t}
	\!-\!\! \( \!\! \frac{\hd (c_s \!+\! \hd \!+\! 1) }{2r^2}
			\!-\! \frac{\hd(\hd \!+\! 1)}{2r} \d_r
			\!-\! \frac{\hd\!+\!1}{2} \w^2
		\! \) \!\! h_{S}
	\!-\!\frac{\hd c_s \hc_s \tilde{h}_{S}}{2(\hd\!+\!1)r^2}
	\nonumber\\
B_3 &:=&
	\( r \d_r - 1 \) \! h_{tt}
	\!+\! i\w \! \( r \d_r - 2 \) \! h_{t}
	\!+\! \frac{\hd}{r} \! \( \frac{2}{r} - \d_r \! \) \!\! h_{S}
	\!-\! r \d_r \! \( \! \frac{\hd\hc_s}{(\hd\!\!+\!\!1)r^2} \tilde{h}_{S} \! \)~.
\eea
We solve these simultaneously to find
\bea
&&~~~~~~~~~~~~~~~~~~~~
\(\begin{matrix}
	h_{tr}\\
	h_{rr}\\
	\frac{c_s}{r}h_{r}
 \end{matrix}\)
= A^{-1} B
= \frac{1}{2 \, \Ms(r)} \,\tilde{A}^{-1} B,
\label{Scalar Algebraic Fields}
\\
\tilde{A}^{-1} &=&
\(\begin{matrix}
  \hd \( \! 2\hd \hc_s \!-\! (\hd\!+\!1) \w^2r^2 \!\)						&
	 ~2i\w r \! \( \! \hd (c_s\!-\!2\hd\!-\!2) \!-\! (\hd\!+\!1) \w^2r^2\!\)		&
	 	-i \w r c_s \hd (\hd\!+\!1)										\\
  *																	&
	-4 c_s\hd											 			&
	\!\!\!\! 2 c_s \( \! (\hd\!+\!1) \w^2r^2\!-\!\hd c_s \! \)	 				\\
  *  																	&
	* 																&
	\!\!\!\!\!\!\!\!\!\!\!\!\! c_s (\hd\!+\!1) \! \( \! 2 (\hd\!+\!1) \w^2 r^2 \!-\! c_s \hd \)
\end{matrix} \)\!,	
\nonumber\\ \nonumber\\
&&~~~~~~~~~~~~~~~~
\Ms(r) := c_s\hc_s\hd^2 - 2\hc_s\hd (\hd\!+\!1)\w^2 r^2 + (\hd\!+\!1)^2 \w^4 r^4, \nonumber
\eea
where the matrix terms under the main diagonal (in asterisks) are completed by hermiticity.
Plugging (\ref{Scalar Algebraic Fields}) back into the action (\ref{homogenous action}-\ref{Action decomposition}), we find that $h_{S}$ drops out entirely, and the three remaining fields $h_{tt}$, $h_{t}$ and $\tilde{h}_{S}$ appear only combined in the single gauge invariant field (compare (4.30) of \cite{AsninKol} in the flat space limit)
\be
\Phi := h_{tt} + 2i\w h_{t} - \w^2 \tilde{h}_{S}~.
\label{Scalar Field}
\ee
The scalar action is thus given by
\bea
\tilde{L}_{S}&=& \frac{1}{\Ms(r)}
	\[	\al \d_r \Phi^{*} \d_r \Phi
		+ \bt \( \d_r \Phi^{*} \Phi
			\!+\! \Phi^{*} \d_r \Phi \)
		+ \gm \Phi^{*} \Phi
	\],~~~~~	\\
\al&=&-\hd(\hd\!+\!1)c_s\hc_s,~
	\bt=\frac{c_s\hc_s}{r}\( c_s \hd \!-\! (\hd\!+\!1) \w^2r^2 \),~
	\gm=\frac{c_s\hc_s}{r^2}\( 2 (\hd\!+\!1) \w^2r^2 \!-\! c_s \hd \).~~~~~~~~~
\eea
Following \cite{AsninKol} we canonically transform to $\tilde{\Phi}$ with the generating function,
\be
F\[\Phi,\Phi^{*},\tilde{\Phi},\tilde{\Phi}^{*}\]
	= - \hc_s r^{\frac{\hd-1}{2}} \( \tilde{\Phi}^{*} \Phi + \tilde{\Phi} \Phi^{*}\)~~,~~
\label{Generating Function}
\ee
and after partial integrations find
\be
L_{S} = \ShPreFCanon \frac{\hd(\hd\!+\!1)\hc_s}{c_s} \tilde{\Phi}^{*}
	\[ \d_r^2 + \w^2 + \(c_s + \frac{\hd^2-1}{4}\) \frac{1}{r^2} \] \tilde{\Phi},
\label{Scalar Action Homegenous Zerilli}
\ee
which we recognize as the Zerilli action \cite{Zerilli:1970se}.
Defining the field
\be
\fieldh = -\frac{\hd (\ell\!+\!\hd\!+\!1)}{2\ell} r^{-(\ell+\frac{\hd\!+\!1}{2})} \tilde{\Phi},
\label{field hS homogenous}
\ee
we find the homogenous scalar action in the familiar master form (compare (\ref{Tensor Action Master Homogenous}, \ref{Vector Action Master Homogenous}))
\be
L_{S} = \ShPreFCanon \frac{4 (\hd\!+\!1) (\ell\!-\!1) \ell}{\hd (\ell\!+\!\hd\!+\!1) (\ell\!+\!\hd)}
	r^{2\ell+\hd+1} \fieldh^{*} \, \IL \, \fieldh \, \, .
\label{Scalar Action Master Homogenous}
\ee

\section{Action in spherical fields: Source terms}
\label{app:Inhomogenous action in spherical fields - Source terms}
We wish to construct the inhomogeneous part of the action $\Sih$ (\ref{action inhomogenous}), which describes the interaction of the gravitational field with matter sources, in terms of the gauge invariant spherical fields and corresponding source functions.
We decompose the stress-energy tensor as
\bea
T^{tt} &=& \sumint T^{tt}_{L\w} n_L e^{-i\w t}   ~~,~~   T^{tr} = \sumint T^{tr}_{L\w} n_L e^{-i\w t} ~~,~~
T^{rr} = \sumint T^{rr}_{L\w} n_L e^{-i\w t} \nonumber\\
T^{t\Om} &=& \sumint \( T^{t}_{L\w} \d^\Om n_L + T^{t\Valph}_{L \w} n_{L}^\AlOm \) e^{-i\w t} ~~,~~
T^{r\Om} = \sumint \( T^{r}_{L\w} \d^\Om n_L + T^{r\Valph}_{L \w} n_{L}^\AlOm \) e^{-i\w t} \nonumber\\
T^{\Om\Om'} &=& \sumint \[ T_{L\w}^{S} n_L^{\Om\Om'}
	+ \tilde{T}_{L\w}^S \tilde{n}_L^{\Om\Om'}
	+ T_{L\w}^\Valph n_{L}^{\AlOm\Om'}
	+ T_{L\w}^{\Talbt} n_{L}^{\albt\Om\Om'} \] e^{-i\w t} \, \, .
\label{decomposition of T}
\eea
We shall also use the inverse transformations,
\bea
T^{(tt/tr/rr)}_{L\w} (r) &=&
	\( \! \Nld \Omd  \) ^{\!\!-1}\!\!\!\!\int\!\!\!\!\int\!\!\dOmd dt e^{i \w t} n_L T^{(tt/tr/rr)} (\vec{r},t)
\label{invTtt}\label{invTtr}\label{invTrr},~~~~~~~\\
T_{L\w} (r) &=&
	\( \! \Nld \Omd  (\hd\!+\!1) \, r^2 \) ^{\!\!-1}\!\!\!\!\int\!\!\!\!\int\!\!\dOmd dt e^{i \w t} n_L \[ T^{aa} (\vec{r},t) -T^{rr} (\vec{r},t) \]
\label{invTS},\\
T^{(t/r)\Valph}_{L\w} (r) &=&
	\( c_s \Nld \Omd \) ^{\!\!-1}\!\!\!\!\int\!\!\!\!\int\!\!\dOmd dt e^{i \w t}
		\eps^{(D)}_{\aleph \, a \, b \, k_\ell} \frac{\ell}{r} \, n^{bL-1} T^{(t/r)a}
\label{invTtV}\label{invTrV},\\
T^{\Talbt}_{L\w} (r) &=& \frac{1}{\Nld \Omd \TensorNorm}
	\!\!\int\!\!\!\!\int\!\!\dOmd dt e^{i \w t}
	\eps^{(D)}_{\aleph a b k_{\ell}} \eps^{(D)}_{\beth a' b' k_{\ell'}}
	\frac{\ell(\ell\!-\!1)}{r^2} n^{bb' \! L-2}  T^{aa'}\!,\,~~~~~
\label{invTT}
\eea
where $T^{\mu r}=T^{r\mu}=\frac{x^c}{r} T^{\mu c}$ and $T^{rr}=\frac{x^c x^d}{r^2} T^{cd}$.

Using the divergence expressions (\ref{spherical divergence}-\ref{spherical divergence Om}), current conservation $D_\mu T^{\mu\nu}=0$ is recast as the $d$ equations (we henceforth suppress $L$, $\w$ indices)
\bea
0&=& - i\w T^{tt}
	+\( \d_r + \frac{\hd+1}{r} \) \!\! T^{tr}
	- c_s T^{t} \, ,
\label{current conservation t}\\
0&=& - i\w T^{tr} + \(\d_r + \frac{\hd+1}{r} \) T^{rr}
	- c_s T^{r} - (\hd+1) r \, T^{S}
\label{current conservation r}\\
0 &=& - i\w T^{t}
	+\(\d_r+\frac{\hd\!+\!3}{r}\) \! T^{r}
	+ T^{S}
	-\frac{\hd \, \hc_s}{(\hd\!+\!1) } \tilde{T}^{S} \, ,
\label{current conservation Om scalar}\\
0 &=&	- i\w T^{t\Valph}
	+\(\d_r+\frac{\hd\!+\!3}{r}\) \! T^{r\Valph}
	- \frac{\hc_s}{2} T^\Valph
\label{current conservation Om vector} \, .
\eea
From these we can eliminate the fields $T^{t}$,$T^{r}$,$\tilde{T}^{S}$, and $T^{\Valph}$, retaining only the scalars $T^{tt}$,$T^{tr}$,$T^{rr}$,$T^{S}$, the vectors $T^{t\Valph}$,$T^{r\Valph}$ and the tensor $T^{\Talbt}$:
\bea
T^{t} &=& \frac{1}{c_s}
	\[ - i\w T^{tt}
	  +\( \d_r + \frac{\hd+1}{r} \) \!\! T^{tr} \],
\label{eliminate Tts}\\
T^{r} &=& \frac{1}{c_s} \[ - i\w T^{tr} + \(\d_r + \frac{\hd+1}{r} \) T^{rr}
							- (\hd+1) r \, T^{S} \],
\label{eliminate TrS}\\
\tilde{T}^{S}
\VERBOSE{	
&=& \frac{\hd\!+\!1}{\hd \, \hc_s}
	\[ - i\w T^{t}
	  +\(\d_r+\frac{\hd\!+\!3}{r}\) \! T^{r}
	  + T^{S} \]	\nonumber\\
}	
&=&\frac{(\hd\!+\!1)}{\hd \, c_s \, \hc_s}
	\[	\( c_s - \(\hd\!+\!1\)\(r \d_r\!+\!\hd\!+\!4\) \) T^{S}
		- \w^2 T^{tt}
		  - 2i\w \( \d_r + \frac{\hd\!+\!2}{r} \) \! T^{tr}
	\right. \nonumber\\ && \left.~~~~~~~~~~
		+\(\d_r^2+2\frac{\hd\!+\!2}{r} \d_r + \frac{(\hd\!+\!1)(\hd\!+\!2)}{r^2}\) T^{rr}
	 \],
\label{eliminate tildeTS}\\
T^{\Valph} &=& \frac{2}{\hc_s}
	\[ - i\w T^{t\Valph}
	  +\(\d_r+\frac{\hd\!+\!3}{r}\) \! T^{r\Valph} \]
\label{eliminate TV}.
\eea

Plugging the expansions (\ref{decomposition of GR fields},\ref{decomposition of T}) into (\ref{action inhomogenous}) and using (\ref{current conservation t}-\ref{current conservation Om vector}) and (\ref{homogenous hrV},\ref{Scalar Algebraic Fields}), we find
\bea
\Sih = -\frac{1}{2} ~ \sumint \int  \[ \fieldh \sourceT^{*} + \fieldh_{\Valph} \, \sourceT^{\Valph \, \, *} + \fieldh_{\Talbt} \, \sourceT^{\Talbt~*} + c.c. \] dr ~ ,
\label{inhomogenous source}
\eea
where the source functions (Scalar, Vector, Tensor) are given by
\bea
\sourceT &:=& -\frac{\Nld \, \Omd \, (\hd+1) \, r^{\ell+\hd+3}}{\hd (\ell\!+\!\hd) (\ell\!+\!\hd\!+\!1)}
	\[ \frac{\hd}{r}\d_r + \w^2 - \frac{\hd\( c_s -2(\hd+1) \)}{(\hd+1) r^2} \] J \, \, ,
\label{TS} \\
J &:=&
	T^{tt}
	+\frac{2i\w}{r^{\hd+1}}
		\left( \frac{(\hd\!+\!1)(-\hd \hat{c}_s + (\hd\!+\!1)r^2 \w^2) r^{\hd+3}}{\Ms} T^{tr} \)'
	+\frac{2i(\hd\!+\!1)^2 r^3 \w^3}{\Ms} T^{tr}
\nonumber \\
&&	-(\hd\!+\!1)^2\[ \frac{1}{r^{\hd+1}}
	\left(  \frac{\w^2 r^{\hd+5}}{\Ms} T^{rr} \)' \, \]'
\nonumber \\
&&	- (\hd\!+\!1)^2 \( \frac{ r^3 \w^2 (c_s \hat{c}_s (\hd\!-\!1)\hd^2
			-2 \hd (\hd\!+\!1)^2 \hat{c}_s r^2 \w^2
			+ (\hd\!+\!3) (\hd\!+\!1)^2 r^4 \w^4)
		} {\Ms^{2}} T^{rr} \right)'
\nonumber \\
&&	+\frac{1}{\Ms^{2}}
		\[c_s^2 \hc_s^2 \hd^3
			\!-\! c_s \hc_s \hd^2 (\hd\!+\!1)(2c_s \!+\! (\hd\!+\!1)(\hd^2\!+\!\hd\!-\!4))r^2 \w^2
		\right.
\nonumber \\
&&~~~~~~~~~\left.
			+ \hc_s \hd (\hd\!+\!1)^2 (c_s\!+\!2\hd(\hd\!+\!1)(\hd\!+\!3))r^4 \w^4
			- (\hd\!+\!1)^4 (\hd\!+\!2)(\hd\!+\!3) r^6 \w^6
		\] T^{rr} ~~~~\nonumber \\
&&	+\frac{(\hd\!+\!1)^3}{r^{\hd+2}}
		\( \frac{\w^2 \, r^{\hd+7}}{\Ms} T^{S} \)'
	+\frac{(\hd\!+\!1) \hc_s \, r^2 (c_s \hd - (\hd\!+\!1) \w^2 r^2)}{\Ms} T^{S}	~ ,
\label{TS2}	\\
\sourceT^{\aleph} &:=&
	- \frac{2 \, \Nld \, \Omd \, (\ell\!+\!\hd) \, r^{\ell+\hd+1}}{\ell\!+\!\hd\!+\!1}
	\left[	r^2 \, T^{t \Valph}
			+\frac{1}{r^{\hd+1}}\left( \frac{i \w r^{\hd+5}}{\Mv} T^{r \Valph} \right)' \,
	\right]' ~ ,
\label{TV}	\\
\sourceT^{\Talbt} &:=&
	\frac{\Nld \, \Omd}{2} \, \hd^2 c_s (c_s - \hd) \, r^{\ell+\hd+3} \, \, T^{\aleph \beth} ~.
\label{TT}
\eea

\section{Origin-normalized Bessel functions}
\label{app:Normalizations of Bessel functions}
For $B$ a Bessel function of the first or second kind or a Hankel function, i.e. $B \in \{J,Y,H^\pm\}$, and with $\al$ representing its order, we define the origin normalized Bessel functions $\tilde{b}_\al$ as
\be
\tilde{b}_\alpha := \Gm(\alpha+1) 2^{\alpha} \frac{B_\alpha(x)}{x^\alpha}\, \, ,
\ee
These functions satisfy the equation (compare \ref{Master Wave Equation})
\be
\[ \del_x^2 + \frac{2\alpha+1}{x}\del_x + 1\] \tilde{b}_\alpha(x) =0 ~.
\label{Modified Bessel equation2}
\ee
The purpose of the definition is to have $\tilde{j}_\alpha$ normalized to $1$ at the origin $x=0$. Around the origin it is given around by the Taylor expansion (which contains only even powers)
\be
\tilde{j}_\alpha (x)  = \sum_{p=0}^\infty \frac{(-)^p \, (2\alpha)!!}{(2p)!! (2p+2\alpha)!!} x^{2p}
= 1 - \frac{x^2}{2(2\al+2)} + \cdots ~.
\label{Bessel J series2}
\ee
The asymptotic form for $x\to\infty$ is best stated in terms of the Hankel functions $\tilde{h}^\pm:= \tilde{j} \pm i \tilde{y}$
\be
\tilde{h}^\pm_\alpha(x) \sim (\mp i)^{\alpha+1/2} \frac{2^{\alpha+1/2} \Gm(\alpha+1)}{\sqrt{\pi}} \frac{e^{\pm i x}}{x^{\alpha+1/2}} \, \, .
\label{Bessel H asymptotic2}
\ee
For more details, see appendix B.2 of \cite{PaperIII}.



\end{document}